\documentclass[iop,apj,caption]{emulateapj}
\usepackage{apjfonts} %If missing fonts, comment out this or google apjfonts to download
\usepackage{amsmath,amstext,mathtools}
\usepackage[breaklinks,colorlinks,citecolor=blue,linkcolor=magenta]{hyperref} 
\usepackage[all]{hypcap} %Links go to figures. This breaks deluxetables; use \capstartfalse \capstarttrue around deluxetables to fix it.
\usepackage{relsize}

\def\red#1 {\textcolor{red}{#1}\ }   %using this command will put the text in red, so as to be easily seen.
\def\blue#1 {\textcolor{blue}{#1}\ }   %using this command will put the text in blue, so as to be easily seen.

\newcommand{\dott}[1]{\skew{3.0}\dot{#1}}

\newcommand{\subphi}{_{\!\!\mathmakebox[0.4em][l]{_\phi}}}
\newcommand{\vel}{\upsilon}
\newcommand{\lcirc}{0.68}
\newcommand{\lecc}{0.81}

\shorttitle{Torque Balance in Circumbinary Disks}
\shortauthors{Mu\~noz, Miranda \& Lai}

\begin{document}

\title{Hydrodynamics of circumbinary accretion: angular momentum transfer and binary orbital evolution}
\author{Diego J. Mu\~noz$^{1,2,3}$, Ryan Miranda$^{3,4}$ \& Dong Lai$^{3}$}
\affil{$^1$
Center for Interdisciplinary Exploration and Research in Astrophysics, Physics and Astronomy,
Northwestern University, Evanston, IL 60208, USA\\
$^2$ Steward Observatory, University of Arizona, Tucson, AZ 85721, USA\\
$^3$Cornell Center for Astrophysics and Planetary Science, Department of Astronomy, Cornell University, Ithaca, NY 14853, USA\\
$^4$Institute for Advanced Study, School of Natural Sciences, Einstein Drive, Princeton, NJ 08540, USA
}
\email[Contact: ]{diego.munoz@northwestern.edu}

\begin{abstract}
We carry out 2D viscous hydrodynamical simulations of circumbinary
accretion using the moving-mesh code {\footnotesize AREPO}. We
self-consistently compute the accretion flow over a wide range of
spatial scales, from the circumbinary disk (CBD) far from the central
binary, through accretion streamers, to the disks around individual
binary components, resolving the flow down to $2\%$ of the binary
separation. We focus on equal-mass binaries with arbitrary
eccentricities.  We evolve the flow over long (viscous) timescales
until a quasi-steady state is reached, in which the mass supply rate at large distances $\dott{M}_0$ (assumed constant) equals the time-averaged
mass transfer rate across the disk and the total mass accretion rate
onto the binary components. 
This quasi-steady state allows us to compute 
the secular angular momentum transfer rate onto the
binary, $\langle \dott{J}_{\rm b}\rangle$, and the resulting orbital
evolution. Through direct computation of the gravitational and
accretional torques on the binary, we find that $\langle \dott
J_{\rm b}\rangle$ is consistently positive (i.e., the binary gains angular
momentum), with $l_0\equiv \langle \dott{J}_{\rm b}\rangle/\dott{M}_0$ in the range
of $(0.4-0.8)a_{\rm b}^2\Omega_{\rm b}$, depending on the binary eccentricity
(where $a_{\rm b},~\Omega_{\rm b}$ are the binary semi-major axis and angular
frequency); we also find that this $\langle \dott{J}_{\rm b}\rangle$ is equal
to the net angular momentum current across the CBD,
indicating that global angular momentum balance is achieved in our
simulations.  In addition, we compute the time-averaged rate of change
of the binary orbital energy for eccentric binaries, and thus obtain
the secular rates $\langle \dot a_{\rm b}\rangle$ and $\langle \dot
e_{\rm b}\rangle$. In all cases, $\langle \dot a_{\rm b}\rangle$ is positive,
i.e., the binary expands while accreting. We discuss the implications
of our results for the merger of supermassive binary black holes and
for the formation of close stellar binaries.
\end{abstract}

\maketitle

\section{Introduction}\label{sec:introduction}
Circumbinary accretion disks are
a natural byproduct of binary star formation via disk fragmentation \citep[e.g.,][]{bos86,bon94a,bon94b,kra08}.
A number of these disks have been observed around Class I/II young stellar binaries -- well known objects include
 GG Tau, DQ Tau and UZ Tau E \citep[e.g.][]{dut94,mat96,mat97} --
and recently even around much younger binary Class 0 objects like L1448 IRS3B \citep{tob16}.
Circumbinary disks are also expected to
form around massive binary black holes following galaxy mergers \citep[e.g.][]{beg80,mil01,mil05,esc05,may07a,dot07,cua09,cha13}, 
and even from the fall-back ejecta around  post-common envelope binaries \citep[e.g.,][]{kas11}.
The interaction between the two
central bodies and the surrounding gas via accretion and gravitational torques dictates the long-term evolution of the binary orbit.
 Thus, understanding the details of binary-disk interaction is essential to explaining the end-state distribution of stellar binary properties, and 
to deciphering how, or if, a circumbinary disk (CBD) can assist the orbital decay of black hole binaries, initiating the process toward merger.

 It is commonly assumed that the binary inevitably loses specific angular momentum 
 by interacting with the outer disk/ring \citep[e.g.,][]{pri91}, which
 results in the migration/coalescence of the binary toward smaller orbital separations. This reasoning
 follows from extending the classical theory of satellite-disk interaction for extreme mass ratios
 \citep[e.g.,][]{lin79,gol80,lin86b,war97}, into the regime of binaries (mass ratios of order unity): the binary and
disk are coupled via Lindblad resonance torques, and
since the CBD rotates at an angular frequency that is lower than the binary's,
 the nonaxisymmetric perturbations exerted in the gas
 ``lag'' the binary, thus exerting a negative torque on it.  Numerous studies \citep[e.g.,][]{gou00,arm02,arm05,cua09,hai09,cha10,far15,raf16,kel17}  
 have modeled the extraction of angular momentum from the binary by the outer disk, suggesting that a ``gas-assisted'' orbital migration 
may be important in facilitating binary black hole mergers at separations that are too wide for gravitational wave radiation 
to  remove orbital angular momentum efficiently (what is known as the ``final parsec'' problem, \citealp{mil03a,mil03b}).

Two complications can alter this picture of binary migration. The first, and most obvious, is that binaries can accrete from the CBD, and this must be taken into account
when computing the torques exerted on the binary by the disk. Accretion torques and gravitational torques can be of a similar order of magnitude:
the change in angular momentum by accretion alone is proportional to the mass accretion rate $\dott{M}$, while the gravitational Lindblad torque is proportional
to the inner disk surface density, which is itself proportional to $\dott{M}$ provided that the disk is in a steady state. 
Because of the time-dependent forcing from the binary, it is only possible to reach a ``quasi-steady state,''  in which the time-averaged mass accretion rate onto the binary matches the mass
transfer rate in the CBD. To achieve this quasi-steady state, simulations require a steady mass supply at large distances from the binary and a long integration time.  
Only under these conditions, can it be guaranteed that the (arbitrary) initial conditions have no impact in the torque balance of the system.
Recently, \citet{mun16b} and \citet{mir17} have shown such quasi-steady-state can indeed be achieved,
at least when the disk viscosity is sufficiently large, in which case the CBD 
transfers, on average, the same amount of gas to the binary
as it would to a single object. Thus, the deep cavity carved in the gas
by the gravitational torques can only {\it modulate} \citep[see][]{art96,gun02,mac08,han10,val11,shi12,dor13,far14,shi15}, but not {\it suppress}, the net accretion onto the binary
\citep[although see][for a claim to the contrary at low viscosities]{rag16}.  {Numerical studies have shown
that accretion can be modulated at either $\sim5P_{\rm b}$ or $P_{\rm b}$  (the binary orbital period), and that the transition between these
two regimes depends on the binary mass ratio \citep{dor13,far14} or binary eccentricity, as shown by
 \citet{mun16b} and \citet{mir17}
 \citep[see also][for qualitatively similar results using smoothed particle hydrodynamics]{art96,hay07}.}

The second complication is the unexplored contribution of the ``circum-single'' disks 
(CSDs) to the total gravitational torque.
In the same way that the outer CBD is expected to contribute negatively to the 
torque on the binary, the CSDs are expected to contribute positively \citep{gol80}, with
the total torque consisting of the net balance of the inner and outer Lindblad torques
(which is typically negative for mass ratios of $\lesssim10^{-4}$; \citealp{war86,war97}).
\citet{sav94} computed the non-linear effect of the CSDs on the binary orbit in the large mass ratio 
($\lesssim1$) regime, concluding that the gas loses angular momentum to the binary, which results in
wave-induced accretion onto the primary \citep[see also][]{ost94,mun15,win18}.
 Whether these inner/CSD torques can be of significance, or even compete with the outer CBD torque is
an open, and largely unexplored, question.

In order to compute the full balance of torques, one needs
to ($i$) resolve the gas dynamics well within the Roche lobes of the primary and secondary, and also
 ($ii$) guarantee that the numerical scheme is able to capture the non-linear response of the CSDs to the tidal forcing by the binary companions. 
 These requirements can be met by shock-capturing hydrodynamical simulations at sufficiently high resolution.
Recently,  novel, Godunov-type numerical schemes with good shock-capturing properties -- such as the {\footnotesize DISCO} \citep{duf11} and  {\footnotesize AREPO} \citep{spr10a,pak16} methods --  have been used to compute the joint gas dynamical evolution of CBDs and CSDs down
to small scales over a large number of binary orbits \citep{far14,far15,mun16b,tan17}.
Our simulations using {\footnotesize AREPO} \citep{mun16b} have uncovered several new features, of short-term 
($\sim P_{\rm b}$) and long-term ($\sim 200P_{\rm b}$) variability, in the accretion rate onto eccentric binaries.

In a systematic exploratory work,  \citet{mir17} (hereafter MML17) simulated viscous disks around binaries with mass ratio ${\sim1}$
 and computed the net angular momentum transfer within a CBD as a function of radius in
quasi-steady-state.  They found that the net angular momentum
transfer rate at the disk's edge to be positive (i.e., angular momentum is transferred inward) for various values of the binary eccentricity.
This implies that in quasi-steady-state, the binary {\it gains} angular momentum from the gas, possibly leading
to binary expansion instead of coalescence. However, the work of MML17 is subject to an important limitation: the gas within the 
binary's orbit was not simulated \citep[see also][]{mac08,dor13}. Although capturing a portion of the circumbinary cavity, the simulations of MML17 excluded the high-density region
where the CSDs form, excising this region by imposing a diode-like boundary condition on an annulus of radius similar to the binary separation.
As a consequence, the complex inflow-outflow nature of the gas inside
the circumbinary cavity \citep{mun16b,tan17} was not captured correctly (this type of boundary may even be prone to some numerical artifacts; \citealp{thu17}).
The purpose of the present work is to remove 
this limitation, by self-consistently computing the gas dynamics within the circumbinary cavity down to scales of $2\%$ of the binary separation, while
simultaneously evolving a CBD that extends out to a distance of almost two orders of magnitude larger than the binary separation. 

As in  \citet{mun16b}  (hereafter ML16), we use the moving-mesh code {\footnotesize AREPO} to simulate circumbinary accretion.  We consider CBDs with a steady mass supply far from the binary.  This supply is enforced in the form of a boundary condition (also implemented  by MML17), which
allows for the possibility of a global steady state, in which the time-averaged accretion rate onto the binary equals to the mass supply rate.
We demonstrate through explicit computations, that in quasi-steady-state, the angular momentum transfer rate within
 the CBD equals the total torque (from gravitational forces and accretion) acting on the binary. This provides
 unambiguous evidence that the binary gains angular momentum from the CBD. Furthermore, by tracking the orbital energy and angular momentum of the accreting binary, 
 { we compute the combined orbit-eccentricity evolution of the binary for the first time on secular timescales.}

This paper is organized as follows. In Section~\ref{sec:methods}, we describe our use of moving-mesh hydrodynamics to simulate circumbinary accretion flows,
and discuss how one can compute the various torques acting on the accreting binary. In Section~\ref{sec:results}, we describe and analyze our simulation
results, identifying the similarities and differences between circular and eccentric binaries. In Section~\ref{sec:orbital_evolution},
we describe our calculation of the orbital energy and angular momentum changes, and 
we translate our results into orbital element evolution. Finally, in Section~\ref{sec:discussion} we summarize our results and discuss
the implications of this work for the formation and evolution of stellar binaries and supermassive binary black holes.

%%%%%%%%%%%%%%%%%%%%%%%%%%%%%%%%%%%%%%%%%%%%%%%%%%%%%
\section{Numerical Methods}\label{sec:methods}
%
%%%%%%%%%%%%%%%%%%%%%%%
\subsection{Moving-mesh Hydrodynamics}\label{sec:num_methods}
We carry out hydrodynamical simulations using the Godunov-type, moving-mesh code {\footnotesize AREPO} \citep{spr10a,pak16}
in its Navier-Stokes version \citep{mun13a,mun13b}.  {\footnotesize AREPO} has been implemented for viscous accretion
disk simulations in 2D (\citealp{mun14}; ML16) and 3D \citep{mun15}. In the present work, we carry out circumbinary accretion simulations
closely following the methodology of ML16. In brief, we simulate two-dimensional, Newtonian viscous flow, with
a locally isothermal equation of state $P=\Sigma c_s^2$. 
The gravitational acceleration
due to the binary's potential $\Phi_{\rm b}(\mathbf{r})$
is evaluated at each cell center, and the motion of the moving cells by action of this external potential is integrated
in a leapfrog-like fashion, i.e., gravity and pure hydrodynamics are ``operator-split'' (see \citealp{spr10a} and \citealp{pak16}). The positions of
the primary and secondary, $\mathbf{r}_1(t)$ and $\mathbf{r}_2(t)$ respectively, as well as their velocities $\mathbf{v}_1(t)$ and $\mathbf{v}_2(t)$,
are prescribed functions of time, determined at each $t$ by solving
the Kepler equation given the binary eccentricity $e_{\rm b}$ and the mass ratio $q_{\rm b}=M_2/M_1$ \citep[e.g][]{mur00}.
 Both $e_{\rm b}$ and the binary semi-major axis $a_{\rm b}$  are fixed constants throughout the duration of the simulation. Such a ``prescribed'' binary is chosen over a ``live'' binary to minimize noise or high frequency ``jittering'' of the binary's orbital elements.
The sound speed $c_s(\mathbf{r})$ is a prescribed function of $\mathbf{r}$, and
is fed into an iterative isothermal Riemann solver at each Voronoi cell interface \citep[e.g.,][]{tor09,spr10a}. The temperature of the disk
is set by the global parameter $h_0$, the disk's vertical aspect ratio. Like the sound speed, the kinematic viscosity coefficient $\nu(\mathbf{r})$
is also a function of $\mathbf{r}$, and proportional to a Shakura-Sunyaev coefficient $\alpha$.
The functions $c_s(\mathbf{r})$ and  $\nu(\mathbf{r})$ are constructed such that $c_s^2\approx h_0^2\mathcal{G}(M_1+M_2)/|\mathbf{r}|$
and $\nu\approx \alpha h_0^2\sqrt{\mathcal{G}(M_1+M_2)|\mathbf{r}|}$ when $|\mathbf{r}|\gg a_{\rm b}$, and that
$c_s^2\approx h_0^2\mathcal{G}M_i/|\mathbf{r}-\mathbf{r}_i|$ and $\nu\approx \alpha h_0^2\sqrt{\mathcal{G}M_i |\mathbf{r}-\mathbf{r}_i|}$ ($i=1,2$)
 when $|\mathbf{r}-\mathbf{r}_i|\ll a_{\rm b}$; i.e., both the CBD and the two CSDs behave like traditional $\alpha$-disks when the flow is approximately Keplerian 
(for more details, see ML16).

We allow the gas to be accreted by the individual members of the binary (from here on, we will refer to the binary components as ``stars,'' although the results presented
here are equally applicable to black hole binaries). Accretion is thus carried out by a sink particle algorithm: as gas cells get closer than a critical distance $r_{\rm acc}$
to one of the stars, they are flagged for their contents to be {\it drained} (cells are not eliminated) at the end of the time step. In order to keep the transition 
at $|\mathbf{r}-\mathbf{r}_i|=r_{\rm acc}$
smooth and avoid severe depletion of cells (which can affect the convergence speed of the Voronoi tessellation algorithm), we accrete cells with $|\mathbf{r}-\mathbf{r}_i|\approx0$
more aggressively than those with $|\mathbf{r}-\mathbf{r}_i|\approx r_{\rm acc}$. This is accomplished by using a smooth spline function \citep{spr01} that 
returns the fraction $f$ of mass and linear momentum that is to be drained from each cell (the spline function is 0 at $|\mathbf{r}-\mathbf{r}_i|=r_{\rm acc}$ and $1$ at $|\mathbf{r}-\mathbf{r}_i|=0$).
In practice, we cap $f$ at a maximum value of 0.5 per time-step to avoid excessive cell depletion \citep[e.g., see][]{kru05}. As cells drift toward the ``star,''
subsequent time-steps will continue to drain their mass and momentum, eventually fully emptying them from their original mass content or merging them into other cells.

The distribution of cells in the CBD is initially ``quasi polar'' \citep{mun14}, with a logarithmic spacing in radius; this type of cell allocation is 
roughly preserved throughout the simulation, during which cells are not allowed
to become too distorted, nor to depart too significantly from their initial size. Within the circumbinary cavity, this polar-like resolution criterion smoothly transitions to
one that is mass-based, i.e., a constant mass per cell is sought. The quasi-Lagrangian nature of {\footnotesize AREPO}  tends to preserve the cell mass along flow
trajectories \citep[e.g.][]{vog12}, however, whenever the finite-volume mass fluxes change a cell's mass, a procedure of cell refinements and de-refinements 
will enforce the desired mass resolution \citep{spr10a}. The target mass is set to be $m_{\rm target}=1.2\times10^{-6}\Sigma_0a_{\rm b}^2$,
where we have introduced a natural scaling for the surface density,
\begin{equation}\label{eq:density_scaling}
\Sigma_0 \equiv \frac{\dott{M}_0}{3\pi \alpha h_0^2\Omega_{\rm b}a_{\rm b}^2}~~,
\end{equation}
with $\dott{M}_0$ the prescribed (constant) mass supply rate at the outer boundary (the cylindrical radius measured from the binary center of mass $R=R_{\rm out}=70 a_{\rm b}$).
We enforce the outer boundary conditions at $R=R_{\rm out}$ by fixing the primitive variables to their standard viscous disk values, i.e., the surface density 
$\Sigma\simeq \dot M_0/(3\pi \nu)$, radial velocity $\vel_R\simeq -(3/2)\nu/R$, and the azimuthal velocity $\vel_\phi\simeq R\Omega_{\rm K}$
 (see Eqs.~\ref{eq:density_profile}, \ref{eq:vel_profile}, \ref{eq:radial_velocity1} and \ref{eq:radial_velocity2} below) with the viscosity given by $\nu=\alpha h_0^2R^2\Omega_{\rm K}$.  
We also add a narrow damping zone (between $60 a_{\rm b}$ and $R_{\rm out}$) to minimize oscillations and nonaxisymmetric features \citep{val06,mun14}.
 We note that the constant-mass resolution is not strictly respected because the complex gas dynamics and density contrast of the circumbinary cavity can
lead to excessively large cells in the dilute regions as well as extremely small cells within the dense regions. In order to circumvent these complications, we impose
a maximum cell area of $\delta A_{\rm max}=1.5\times10^{-3}a_{\rm b}^2$ and a minimum of $\delta A_{\rm min}=10^{-5}a_{\rm b}^2$. 
As a rule of thumb, we require $\delta A_{\rm min}$  to be smaller than $\sim\pi r_{\rm acc}^2$ by a factor of at least 100.

The binary properties are fully determined by $e_{\rm b}$ and the mass ratio $q_{\rm b}=M_2/M_1$. In this paper, we only vary
$e_{\rm b}$, fixing $q_{\rm b}=1$. We work in distance units relative to the binary semimajor axis, so $a_{\rm b}=1$. Similarly,
we set $\mathcal{G}=1$ and $M_{\rm b}=M_1+M_2=1$. As noted above, the disk
properties are set by the viscosity parameter $\alpha$ and the dimensionless aspect ratio $h_0$. Throughout this work, we fix $\alpha=h_0=0.1$. The
fiducial accretion radius is set at $r_{\rm acc}=0.02a_{\rm b}$, which is the same as the softening length of the Keplerian potential of each ``star.''

%%%%%%%%%%%%%%%%%%%%%%%%%%%
\subsection{Torque Acting on an Accreting Binary}
In the following, we describe how we compute the angular momentum change of the binary from simulation output in post-processing.
In particular, we detail how the {\it direct} torque on the binary can be decomposed into all of its relevant contributions (Section~\ref{sec:direct_torques}).
We also explain how
the torque on the binary can be related to the angular momentum transfer as a function of radius in the CBD (Section~\ref{sec:angmom_transfer}).

%%%%%%%%%%%%%%%%%%%%%%%
\subsubsection{Direct Computation of Torque Acting on a Binary}\label{sec:direct_torques}
The orbital angular momentum of the binary is subject to torques due to accretion and gravitational interactions:
\begin{equation}\label{eq:angmom_change}
\begin{split}
\frac{d L_{\rm b}}{dt} & = \left.\frac{d L_{\rm b}}{dt}\right|_{\rm acc} + \left.\frac{d L_{\rm b}}{dt}\right|_{\rm grav}\\
&= \dot{\mu}_{\rm b}l_{\rm b}
+\mu_{\rm b}\frac{d l_{\rm b}}{dt}\Big|_{\rm acc} +\mu_{\rm b}\frac{d l_{\rm b}}{dt}\Big|_{\rm grav}~,
 \end{split}
\end{equation}
where $l_{\rm b}=a_{\rm b}^2\Omega_{\rm b} (1-e_{\rm b}^2)^{1/2}$ is the binary's specific angular momentum, and is defined by $L_{\rm b}=\mu_{\rm b}l_{\rm b}$, where
 $\mu_{\rm b}=M_1M_2/M_{\rm b}=q_{\rm b}M_{\rm b}/(1+q_{\rm b})^2$ is the binary's reduced mass.
For the  low-mass disks of interest in this paper, the quantities $M_{\rm b}$, $q_{\rm b}$ and $l_{\rm b}$ in Equation~(\ref{eq:angmom_change})
change a negligible amount over the timescales of the simulations (i.e., these quantities evolve over a long, secular timescale, $\sim M_{\rm b}/\dott{M}_{\rm b}$), and we determine their time
 derivatives from simulation output in
post-processing. Hence, the binary evolution is not ``live'' (see Section~\ref{sec:num_methods} above).

The  rate of change of {\it total} angular momentum in the binary $\dott{J}_{\rm b}$ 
should also include the angular momentum transfer to the spins of the "stars". Thus
\begin{equation}\label{eq:angmom_change2}
\dott{J}_{\rm b}=\dot{L}_{\rm b} + \dot{S}_1 +\dot{S}_2~.
\end{equation}
Below, we describe in detail how the different contributions to Eqs.~(\ref{eq:angmom_change}) and~(\ref{eq:angmom_change2}) are computed from numerical simulations.

\paragraph{Mass Change}%%%%%%%%%%%%%%%%%%%%%%%%%%%%%%
{Even in the absence of specific torques in Eq.~(\ref{eq:angmom_change}), the angular
momentum of the binary changes via changes in the reduced mass $\mu_{\rm b}$.}
Theses changes are simply related to the changes in mass ratio and total mass:
\begin{equation}\label{eq:reduced_mass_change}
\frac{\dot{\mu}_{\rm b}}{\mu_{\rm b}}=\frac{\dott{M}_{\rm b}}{M_{\rm b}}+\frac{1-{q}_{\rm b}}{1+{q}_{\rm b}}\frac{\dot{q}_{\rm b}}{q_{\rm b}}~,
\end{equation}
with
\begin{equation}\label{eq:mass_change}
\dott{M}_{\rm b} = \dott{M}_1 + \dott{M}_2\;\;\;\text{and}\;\;\;\;\;\;
\dot{q}_{\rm b}  = \frac{1+q_{\rm b}}{M_{\rm b}}\left(\dott{M}_2-q_{\rm b}\dott{M}_1\right)~.
\end{equation}
As in ML16, we compute the accretion rate by keeping track of $\Delta m_{{\rm acc},i}(t)$, the cumulatively accreted mass 
at time $t$,  over
the course of the simulation. We compute $\dott{M}_i$ in post-processing, by taking the time derivative
using a centered-differences approximation:
\begin{equation}\label{eq:accretion_calculation}
\dott{M}_i(t) \equiv \frac{d}{dt} \Delta m_{{\rm acc},i}(t)~.
\end{equation}

\paragraph{Specific gravitational torque}%%%%%%%%%%%%%%%%%%%%%%%%%%%
The specific gravitational torque acting on the binary has the general form:
\begin{equation}\label{eq:torque_grav}
\begin{split}
\frac{d \mathbf{l}_{\rm b}}{dt}\Big|_{\rm grav}&= 
\frac{M_1}{\mu_{\rm b}} \mathbf{r}_1{\times}\mathbf{f}_{\rm grav,1} + 
\frac{M_2}{\mu_{\rm b}} \mathbf{r}_2{\times} \mathbf{f}_{\rm grav,2}
= \mathbf{r}_{\rm b}\times (\mathbf{f}_{\rm grav,1}-\mathbf{f}_{\rm grav,2})~,
\end{split}
\end{equation}
where $\mathbf{r}_{\rm b}=\mathbf{r}_{1}-\mathbf{r}_{2}$,  and $\mathbf{f}_{{\rm grav},i}$ is
 the force per unit mass acting on the primary/secondary:
\begin{equation}
\mathbf{f}_{{\rm grav},i}=-\mathcal{G}\sum_k m_k \frac{\mathbf{r}_1-\mathbf{r}_k}{|\mathbf{r}_1-\mathbf{r}_k|^3}~,
\end{equation}
where $m_k$ and $\mathbf{r}_k$ are the mass and position vector of the $k$-th computational cell and the sum extends over all cells in the
simulation domain. A proper account of the acceleration history of the ``stars'' requires a densely sampled force evaluations in time. In practice,
we follow the gravitational ``kicks'' that act on the primary and secondary,  evaluating them twice
per time-step, mimicking the kick-drift-kick operation that one would implement for a truly ``live'' binary orbit integration. As with the accreted mass,
at each time $t$, we store the cumulative velocity change  $\Delta \mathbf{v}_{{\rm grav},i}(t)$. In post-processing, the gravitational force per unit mass is
computed as
\begin{equation}\label{eq:gravitational_force}
\mathbf{f}_{\rm grav,i}= \frac{d}{dt} \Delta \mathbf{v}_{{\rm grav},i}(t)~.
\end{equation}
This approach is based on momentum conservation; as such, it takes into account all time-steps in the simulation and does not rely on the instantaneous
forces computed from sparse simulation snapshots.

\paragraph{Specific accretion torque}%%%%%%%%%%%%%%%%%%%%%%%%%%%
Accretion itself can induce a specific torque on the binary \citep[e.g., see][]{roe12}, separate from
the torque inherent to the change in binary mass.  
Similar to Equation~(\ref{eq:torque_grav}), we have:
\begin{equation}\label{eq:torque_accretion}
\frac{d \mathbf{l}_{\rm b}}{dt}\Big|_{\rm acc}=\mathbf{r}_{\rm b}\times (\mathbf{f}_{\rm acc,1}-\mathbf{f}_{\rm acc,2})~~.
\end{equation}
We compute the accretion forces by keeping track of the velocity kicks as mass is incorporated into the ``stars''. 
As one would do in ``live'' accretion integrations, conservation of linear momentum dictates that the velocity of an accreting body of mass $M_i$ is updated
after each time-step according to
\begin{subequations}\label{eq:accretion_kick}
\begin{align}
\delta \mathbf{v}_{{\rm acc},i}&=\frac{\mathbf{v}_i(t)\,M_i+\delta\mathbf{p}_{{\rm acc},i}}{M_i+\delta m_{{\rm acc},i}}-\mathbf{v}_i(t)\\
&\approx \frac{\delta\mathbf{p}_{\rm acc,i}}{M_i} - \frac{\delta m_{\rm acc,i}}{M_i}\mathbf{v}_i(t)~~,
\end{align}
\end{subequations}
where $\delta m_{{\rm acc},i}$ and $\delta\mathbf{p}_{{\rm acc},i}$ are the changes in mass and momentum during a single time-step.
The accumulation of $\delta \mathbf{v}_{{\rm acc},i}$ over time then defines $\Delta \mathbf{v}_{{\rm acc},i}(t)$. As in Equation~(\ref{eq:gravitational_force}),
the accretion force is computed in post-processing as
\begin{equation}\label{eq:accretion_force}
\mathbf{f}_{{\rm acc},i}= \frac{d}{dt} \Delta \mathbf{v}_{{\rm acc},i}(t)~~,
\end{equation}
from which we  compute the associated torques (Equation~\ref{eq:torque_accretion}). This specific torque vanishes if  accretion onto
the ``star'' is isotropic, in which case,  only the mass -- but not the velocity -- changes. Therefore, we typically
refer to the torque resulting from Equation~(\ref{eq:accretion_force}) as the anisotropic accretion torque.

Note that, had we stored the cumulative evolution of accreted linear momentum $\Delta \mathbf{p}_{{\rm acc},i}(t)$ instead of $\Delta \mathbf{v}_{{\rm acc},i}(t)$, 
the {\it total} accretion torque could be written as
\begin{equation}\label{eq:accretion_torquef_full}
\begin{split}
\frac{d\mathbf{L}_{\rm b}}{dt}\bigg|_{\rm acc}&=
 \mathbf{r}_1\times \frac{d}{dt} \Delta \mathbf{p}_{{\rm acc},1}(t) + \mathbf{r}_2\times \frac{d}{dt} \Delta \mathbf{p}_{{\rm acc},2}(t)\\
 &\approx  \mu_{\rm b} \mathbf{r}_{\rm b}\times (\mathbf{f}_{{\rm acc},1}-\mathbf{f}_{{\rm acc},2})+(\mathbf{r}_{\rm b}\times\mathbf{v}_{\rm b})\frac{d\mu_{\rm b}}{dt}~,
 \end{split}
\end{equation}
{where $\mathbf{v}_{\rm b}=\dot{\mathbf{r}}_{\rm b}=\mathbf{v}_1-\mathbf{v}_2$. Note that the 
second term in Equation~(\ref{eq:accretion_torquef_full}) is already accounted for by the first term in Equation~(\ref{eq:angmom_change}), which describes
the accretion-induced torque at constant specific angular momentum.}

%%%%%%%%%%%%%%%%%%%%%%%%%%%%%%%%%%
\paragraph{Spin Torque}
Gas cells  typically orbit around the ``stars'' before being drained, therefore carrying a small amount of angular momentum relative to a moving center.
This residual angular momentum is irrevocably lost from the system
due to accretion \citep[e.g.,][]{bat95}. The amount of angular momentum -- or ``spin'' --  $\Delta \mathbf{S}_{{\rm acc},i}$  that is transferred to the ``star'' can be tracked as follows: 
each time some mass and momentum are drained from the $k$-th gas cell, the spin angular momentum $(\mathbf{r}_k-\mathbf{r}_i)\times  f_k \mathbf{p}_k$
(where $f_k$ is the fraction of mass {\it and} linear momentum drained from the $k$-th cell) is added to $\Delta \mathbf{S}_{{\rm acc},i}$\footnote{%%%%%%%%%%%%%%%
{The angular momentum of the accreted portion of a gas cell in barycentric coordinates 
$\mathbf{r}_k\times  f_k\mathbf{p}_k$ can be trivially split into an orbital
part  $\mathbf{r}_i\times  f_k \mathbf{p}_k$,  which is automatically transferred to the binary via accretion
(see Eq.~\ref{eq:accretion_torquef_full}), and the ``spin'' part $(\mathbf{r}_k-\mathbf{r}_i)\times  f \mathbf{p}_k$,
which is lost. More specifically, the angular momentum of the $i$-th ``star'' and surrounding gas before accretion
$\mathbf{J}=M_i\mathbf{r}_i{\times}\mathbf{v}_i+\sum \mathbf{r}_k{\times}\mathbf{p}_k$ (sum over contributing cells)
differs from the one after accretion 
$\mathbf{J}'=M_i'\mathbf{r}_i{\times}\mathbf{v}_i'+\sum \mathbf{r}_k{\times}\mathbf{p}_k'$ by the residual amount $\sum(\mathbf{r}_k-\mathbf{r}_i)\times (\mathbf{p}_k-\mathbf{p}_k')=\sum(\mathbf{r}_k-\mathbf{r}_1)\times f_k\mathbf{p}_k$. Linear momentum and mass are updated as $\mathbf{p}_k'=(1-f_k)\mathbf{p}_k$ and  $m_k'=(1-f_k)m_k$  for each accretion event, where $M_i\gg m_k$ is assumed.}
}. %%%%%%%%%%%%%%%%%%%%%%%%%%%%%%%%%%%%%%%%%%%%%%%%%%%%%%%%%%%%%%%%%%%%%%%%%%%%
As before, the spin torque is computed in post-processing as:
\begin{equation}\label{eq:torque_spin}
\dot{\mathbf{S}}_{{\rm acc},i}= \frac{d}{dt} \Delta \mathbf{S}_{{\rm acc},i}(t)~~.
\end{equation}

In summary, in order to compute $\dott{L}_{\rm b}$ and $\dott{J}_{\rm b}$
(Eqs.~\ref{eq:angmom_change} and~\ref{eq:angmom_change2}) from simulation output, we 
need: $\dott{M}_{\rm b}$ and $\dot{q}_{\rm b}$ from Eqs.~(\ref{eq:mass_change}) and using
Equation~(\ref{eq:accretion_calculation});  $({d l_{\rm b}}/{dt})\big|_{\rm grav}$  from Equation~(\ref{eq:torque_grav});
 $({d l_{\rm b}}/{dt})\big|_{\rm acc}$  from Equation~(\ref{eq:torque_accretion}); and $\dott{S}_{1,2}$ from Equation~(\ref{eq:torque_spin}).

To determine the secular orbital evolution of an eccentric binary, we also need to compute the rate of change of
the orbital energy, which is fully determined from $\dott{M}_{\rm b}$, $\mathbf{f}_{\rm grav,i}$ and
$\mathbf{f}_{\rm acc,i}$. This is discussed in Section~\ref{sec:orbital_evolution}.

%%%%%%%%%%%%%%%%%%%%%%%%%%%%%%%%%%%%%%%%%%%%%%%%%%
\subsubsection{Angular Momentum Transfer in the Circumbinary Disk}\label{sec:angmom_transfer}
In quasi-steady-state, the time-averaged angular momentum transfer rate across the CBD,
$\langle\dott{J}_{\rm d}(R,t)\rangle$, is constant in $R$ and equal to
the angular momentum transfer rate onto the binary $\langle\dott{J}_{\rm b}\rangle$. The balance
of angular momentum currents throughout the CBD is (see appendix~A of MML17):
\begin{equation}\label{eq:angular_momentum_balance}
\dott{J}_{\rm d} = \dott{J}_{\rm d,adv}  -  \dott{J}_{\rm dvisc} - \dott{J}_{\rm d,grav}~,
\end{equation}
where  $\dott{J}_{\rm d,adv}$, $\dott{J}_{\rm d,visc}$ and  $\dott{J}_{\rm d,grav}$ are the advective, viscous and gravitational
contributions respectively (in the notation of MML17  $\dott{J}_{\rm d,grav}$ is $T_{\rm grav}^{>r}$).
 The computation of the angular momentum current in the disk $\dott{J}_{\rm d}$ from the {\footnotesize AREPO} output
is more involved than in the {\footnotesize PLUTO} case, since there is no polar grid. In order to
obtain $\dott{J}_{\rm d,adv}$, $\dott{J}_{\rm d,visc}$ and  $\dott{J}_{\rm d,grav}$, we first compute 2D
maps of angular momentum fluxes $F_J(R,\phi)$ by remapping/interpolating the primitive variables of {\footnotesize AREPO} output
onto a structured polar grid. The currents $\dott{J}(R)$ are then calculated from
\begin{equation}\label{eq:angular_momentum_transfer}
\dott{J}(R) = \int_0^{2\pi} F_J(R,\phi) Rd\phi = 2\pi \langle F_J(R,\phi) R\rangle\subphi~~.
\end{equation}
The advective, viscous and gravitational fluxes are, respectively, 
\begin{align}
\label{eq:flux_adv}
F_{J,{\rm adv}}&= -R  \Sigma \vel_R\vel_\phi = -\Sigma  \frac{\left[xy(\vel_y^2-\vel_x^2)+\vel_x \vel_y(x^2-y^2)\right]}{\sqrt{x^2+y^2}}~~,
\\
\label{eq:flux_visc}
F_{J,{\rm visc}} & =-R^2\nu\Sigma\left[\frac{\partial}{\partial R}\left(\frac{\vel_\phi}{R}\right)+\frac{1}{R^2}\frac{\partial \vel_R}{\partial \phi}\right] \\
\nonumber&= - \frac{\nu\Sigma}{\sqrt{x^2+y^2}}
\bigg[
{2xy}\left(\frac{\partial \vel_y}{\partial y}-\frac{\partial \vel_x}{\partial x}\right)
+{(x^2-y^2)}\left(\frac{\partial \vel_x}{\partial y}+\frac{\partial \vel_y}{\partial x}\right)
\bigg]\\
\nonumber\text{and}&\\
\label{eq:flux_grav}
F_{J,{\rm grav}} &=-\int_R^{R_{\rm out}} \Sigma \frac{\partial \Phi_{\rm b}}{\partial \phi} dR'=-\int_R^{R_{\rm out}}  \Sigma \left(-x a_y + y a_x\right) dR'~~,
\end{align}
where we have made the dependence on cartesian coordinates (the ones used by {\footnotesize AREPO}) explicit. In
Equation~(\ref{eq:flux_grav}),  $\mathbf{a}=(a_x,a_y)=-\nabla\Phi_{\rm b}$ is the cell-centered acceleration of the gas due to the binary's potential $\Phi_{\rm b}$.
These maps are computed from simulation snapshots, stacked to produce a time average and then integrated over azimuth (Equation~\ref{eq:angular_momentum_transfer})
to produce the time-averaged angular momentum currents of Equation~(\ref{eq:angular_momentum_balance}).

\subsection{Initial Setup}%%%%%%%%%%%%%
\label{sec:initial_setup}
As in ML16, we prescribe an initially axisymmetric disk model 
(their Eq.~2) with fixed parameters of Shakura-Sunyaev viscosity $\alpha$ and disk aspect ratio $h_0$.
  Motivated by the lessons learned from the numerical exploration of MML17,
 we use a slightly modified
initial condition of the form:
\begin{equation}\label{eq:density_profile}
\Sigma(R)=\Sigma_0 \left(\frac{R}{a_{\rm b}}\right)^{-1/2}\left[1-\frac{l_{0,_{\rm guess}}}{\Omega_{\rm b}a_{\rm b}^2}\sqrt{\frac{a_{\rm b}}{R}}\right]\times\mathcal{F}(R;R_{\rm cav})~,
\end{equation}
with $\Omega_{\rm b}=({\mathcal{G}M_{\rm b}/a_{\rm b}^3})^{1/2}$ and $\Sigma_0$ given by Eq.(~\ref{eq:density_scaling}), 
where ${l}_{0,_{\rm guess}}$ is an approximate initial guess for 
the angular momentum per unit accreted mass ${l}_{0}\equiv{\langle\dott{J}_{\rm b}\rangle}/{\langle\dott{M}_{\rm b}\rangle}$, carried by the accretion
flow from the CBD onto the binary. We adopt the initial guess ${l}_{0,{\rm guess}}=0.75+0.35e_{\rm b}$.
The factor $\mathcal{F}(R;R_{\rm cav})$ in Equation~(\ref{eq:density_profile}) is a tapering function responsible for carving a cavity
in the density profile (it goes to 0 as $R\rightarrow0$ and to 1 for $R\gtrsim2R_{\rm cav}$). We find that a model function of the form
\begin{equation}\label{eq:cavity_profile}
\mathcal{F}(R;R_{\rm cav})=\exp\left[-\left(\frac{R_{\rm cav}}{R}\right)^{4}+\left(\frac{R_{\rm cav}}{R}\right)^{2}\right]
\end{equation}
provides a good fit for $\Sigma$ for all the values of $e_{\rm b}$ explored by MML17 ($q_{\rm b}=1$ fixed).  
The fitted values of $R_{\rm cav}$ increase nearly monotonically
with binary eccentricity: we find  
\begin{equation}\label{eq:cavity_size}
R_{\rm cav}\approx\left\{
\begin{array}{lr}
 2.45 a_{\rm b} & \text{for}\;\;\; e_{\rm b}<0.1\\
 a_{\rm b}(1.5 e_{\rm b} + 2.3) & \text{for}\;\; e_{\rm b}\geq0.1~~.
\end{array}\right.
\end{equation}
The profile (\ref{eq:density_profile}) is discretized into 400 azimuthal
 zones and 600 radial zones logarithmically spaced between $R_{\rm in}=(1+e_{\rm b})a_{\rm b}$ and $R_{\rm out}=70a_{\rm b}$.

As usual, the azimuthal velocity profile is supplied to satisfy centrifugal equilibrium:
\begin{equation}\label{eq:vel_profile}
\vel_\phi^2(R)=\frac{\mathcal{G}M_{\rm b}}{R}\left[1 + 3\frac{Q}{R^2}\right]-c_s^2(R)\left[1 - \frac{R}{\Sigma}\frac{d\Sigma}{dR}\right]~~,
\end{equation}
where the sound speed $c_s^2(R)\propto h_0^2R^{-1}$ is a fixed function of radius alone for $R\gg a_{\rm b}$, and where the quadrupole correction $Q$
 to the binary's (orbit-averaged) gravitational potential is given by:
\begin{equation}
Q=\frac{a_{\rm b}^2}{4}\frac{q_{\rm b}}{(1+q_{\rm b})^2}\left(1+\frac{3}{2}e_{\rm b}^2\right)~~.
\end{equation}
%

 %%%%%%%%%%%%%%%%%%%%%%%%%%%%%%%%%%%%%%%%%%%%%%%%%%%
%%%%%%%%%%%%%%%%%%%%%%%%%%%%%%%%%%%%%%%%%%%%%%%%%%%
\begin{figure}[h!]
\includegraphics[width=0.49\textwidth]{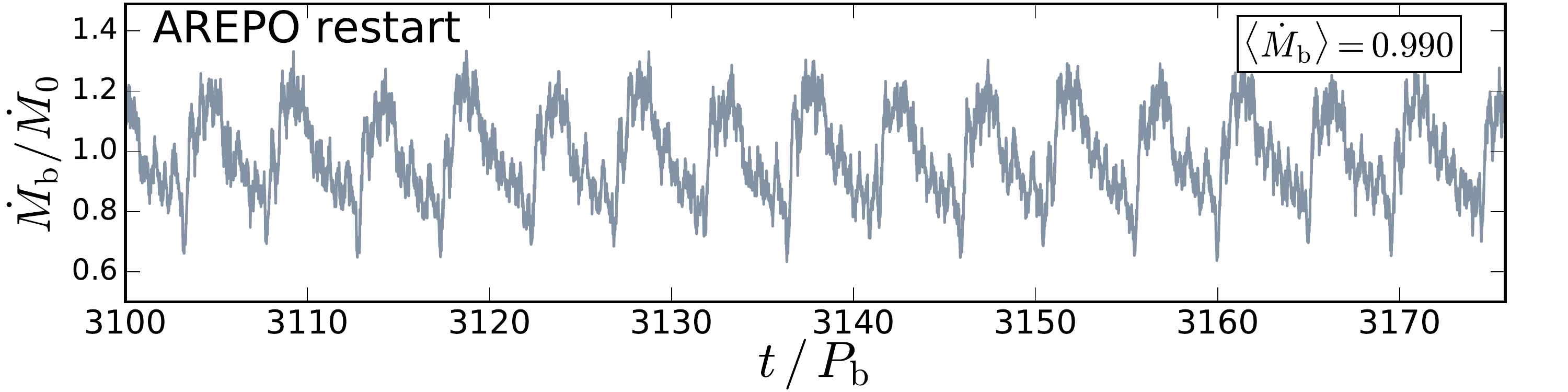}
\includegraphics[width=0.49\textwidth]{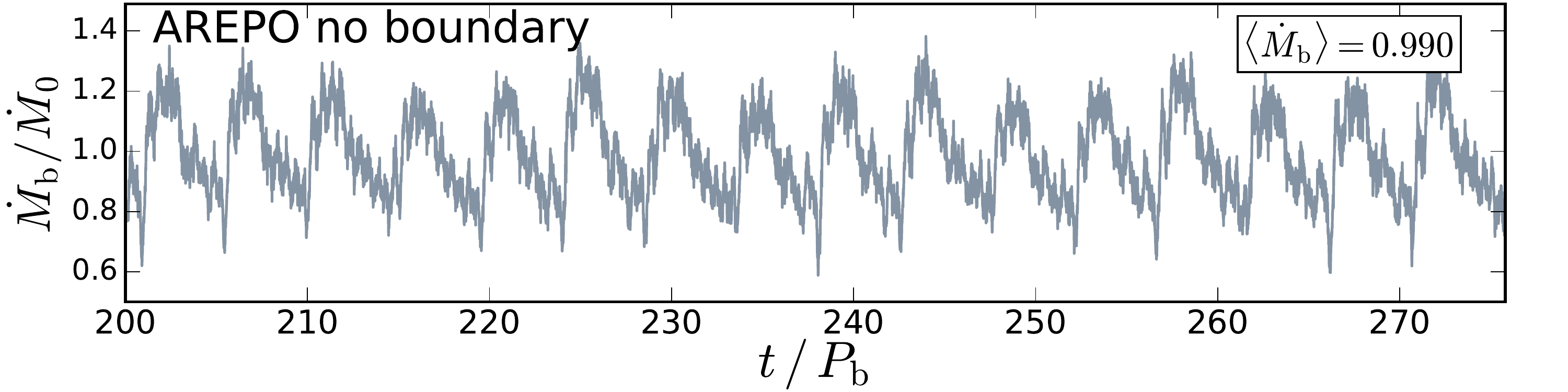}
\includegraphics[width=0.49\textwidth]{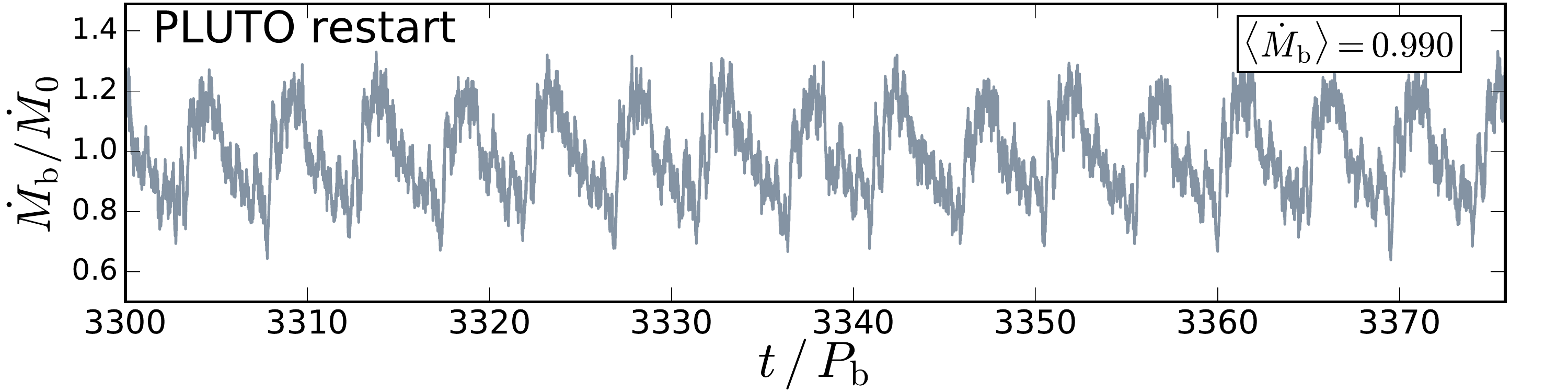}
\caption{Total binary accretion rate $\dott{M}_{\rm b}=\dott{M}_1+\dott{M}_2$ for three steady-state tests over a period of 75.8 orbits
(chosen to cover approximately 16 cycles of the accretion burst with frequency $\sim\tfrac{1}{5}\Omega_{\rm b}$).
The same quasi-steady state is reached regardless of the different initial conditions (see Section~\ref{sec:steady_state}).
\label{fig:accretion_rates}}
\vspace{-0.1in}
\end{figure}
%%%%%%%%%%%%%%%%%%%%%%%%%%%%%%%%%%%%%%%%%%%%%%%%%%%
%%%%%%%%%%%%%%%%%%%%%%%%%%%%%%%%%%%%%%%%%%%%%%%%%%%

%%%%%%%%%%%%%%%%%%%%%%%%%%%%%%%%%%%%%%%%%%%%%%%%%%%
%%%%%%%%%%%%%%%%%%%%%%%%%%%%%%%%%%%%%%%%%%%%%%%%%%%
\begin{figure*}[t!]
\vspace{-0.05in}
\centering
\includegraphics[width=0.83\textwidth]{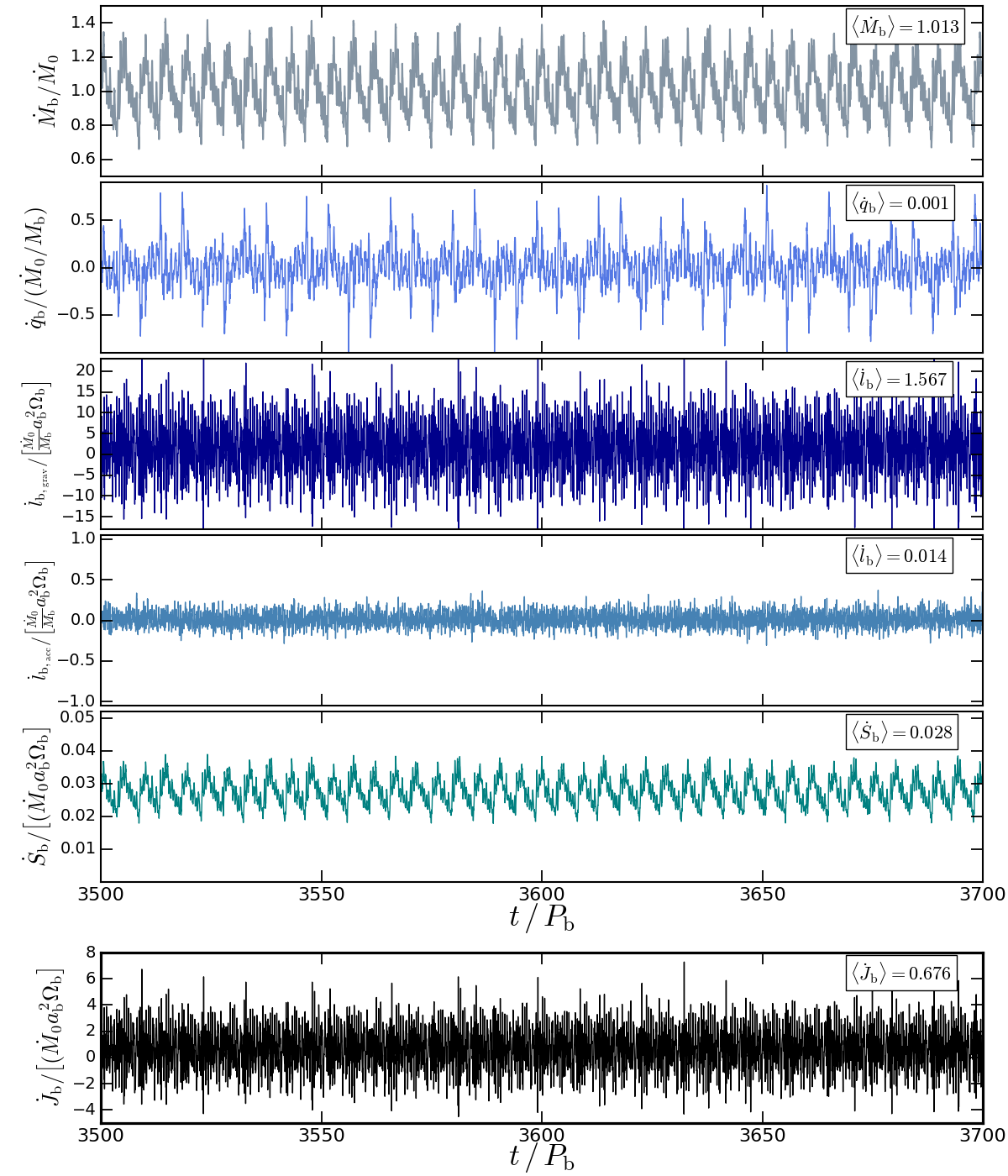}
\vspace{-0.1in}
\caption{The five different contributions to angular momentum transfer rate for an accreting circular binary
and their combined effect $\dott{J}_{\rm b}$ {according to Equations~(\ref{eq:angmom_change}) and~(\ref{eq:reduced_mass_change})}.
 From top to bottom  $\dott{M}_{\rm b}$, $\dot{q}_{\rm b}$, 
$\dot{l}_{\rm b,grav}$, $\dot{l}_{\rm b,acc}$, $\dot{S}_{\rm b}\equiv\dot{S}_1+\dot{S}_2$  (see Section~\ref{sec:direct_torques}); 
and  $\dott{J}_{\rm b}$. In steady state, $\langle\dott{M}_{\rm b}\rangle\approx \dott{M}_0$
and $\langle\dott{J}_{\rm b}\rangle \approx0.676\dott{M}_0\Omega_{\rm b}a_{\rm b}^2$.
Each time series is approximately stationary, with the time-averaged value indicated. 
The time sampling interval in each panel is $\approx0.02P_{\rm b}$.
The accretion eigenvalue in this case
is $l_0\equiv\langle\dott{J}_{\rm b}\rangle/\langle\dott{M}_{\rm b}\rangle\approx\lcirc \Omega_{\rm b}a_{\rm b}^2$~.
\label{fig:angmom_components}}
\end{figure*}
%%%%%%%%%%%%%%%%%%%%%%%%%%%%%%%%%%%%%%%%%%%%%%%%%%%
%%%%%%%%%%%%%%%%%%%%%%%%%%%%%%%%%%%%%%%%%%%%%%%%%%%

%%%%%%%%%%%%%%%%%%%%%%%%%%%%%%%%%%%%%%%%%%%%%%%%%%%
%%%%%%%%%%%%%%%%%%%%%%%%%%%%%%%%%%%%%%%%%%%%%%%%%%%
\begin{figure}[t!]
\centering
\vspace{-0.05in}
\includegraphics[width=0.45\textwidth]{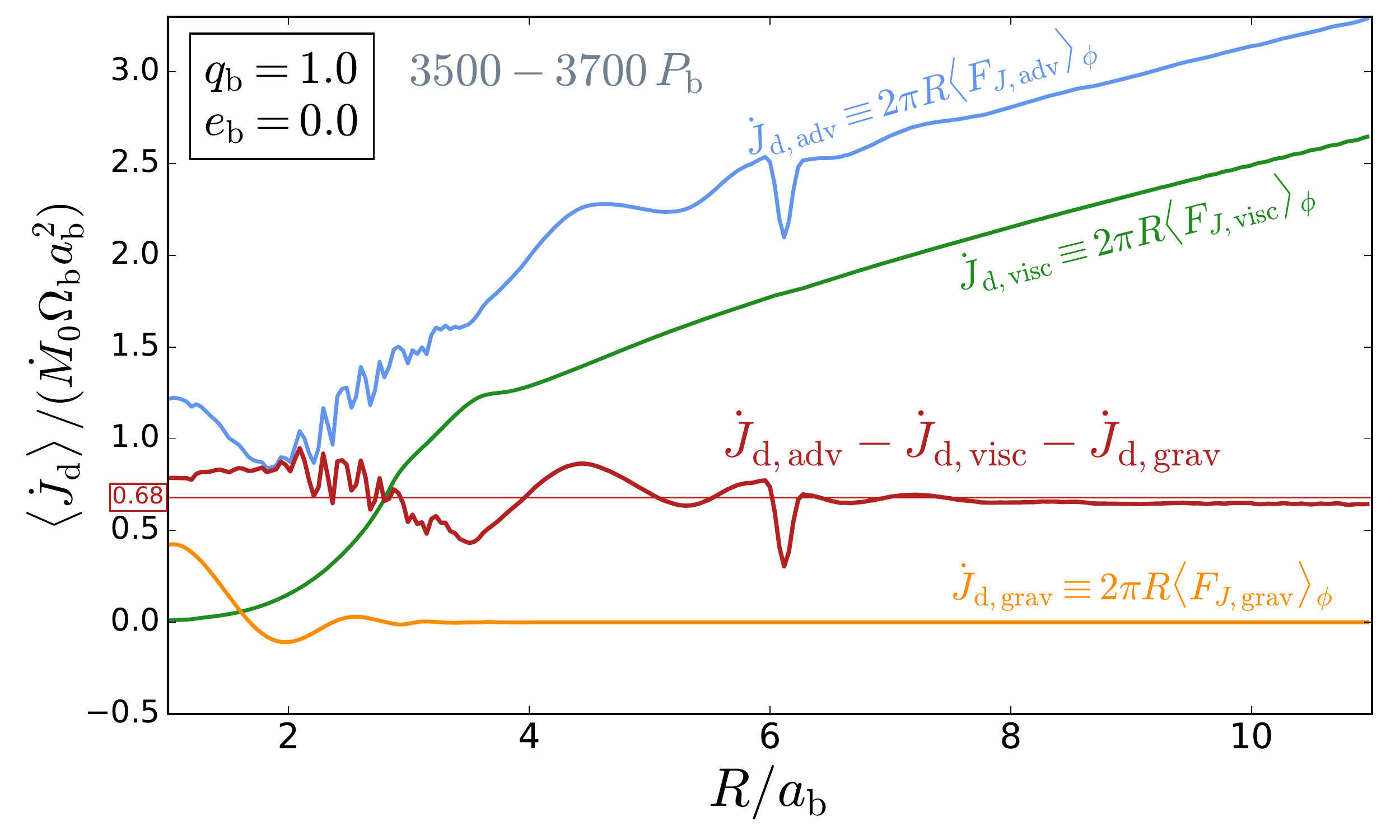}
\vspace{-0.12in}
\caption{Time-averaged angular momentum currents due to advection, viscosity and gravitational torques
 in the CBD for binary parameters $q_{\rm b}=1$ and $e_{\rm b}=0$ (see Section~\ref{sec:angmom_transfer}). The angular momentum currents are computed 
 from the azimuthal integral of the angular momentum flux $F_J$ (Equation~\ref{eq:angular_momentum_transfer}), and  is time-averaged over a
interval of $200\,P_{\rm b}$. The red curve is approximately constant, indicating quasi-steady-state. For reference, the net angular momentum change rate of the binary 
$\langle\dott{J}_{\rm b}\rangle=0.68\dott{M}_0\Omega_{\rm b}a_{\rm b}^2$
is overlaid as a straight red line.
 {The ``blip'' at $R=6a_{\rm b}$ and fluctuations at $R\approx2.5a_{\rm b}$}  are artifacts of the mapping from the Voronoi cells onto a regular polar grid and
their locations are sensitive to resolution in both the original simulation as in the remapped radial bins; 
such irregularities only appear in the $\dot{J}_{\rm adv}$ profile,
which is always the noisiest of the different contributions to $\dott{J}_{\rm d}(R)$. 
\vspace{-0.01in}
\label{fig:angmom_currents}}
\end{figure}
%%%%%%%%%%%%%%%%%%%%%%%%%%%%%%%%%%%%%%%%%%%%%%%%%%%
%%%%%%%%%%%%%%%%%%%%%%%%%%%%%%%%%%%%%%%%%%%%%%%%%%%

%%%%%%%%%%%%%%%%%%%%%%%%%%%%%%%%%%%%%%%%%%%%%%%%%%%
%%%%%%%%%%%%%%%%%%%%%%%%%%%%%%%%%%%%%%%%%%%%%%%%%%%
\begin{figure*}[t!]
\vspace{-0.1in}
\includegraphics[width=0.3\textwidth]{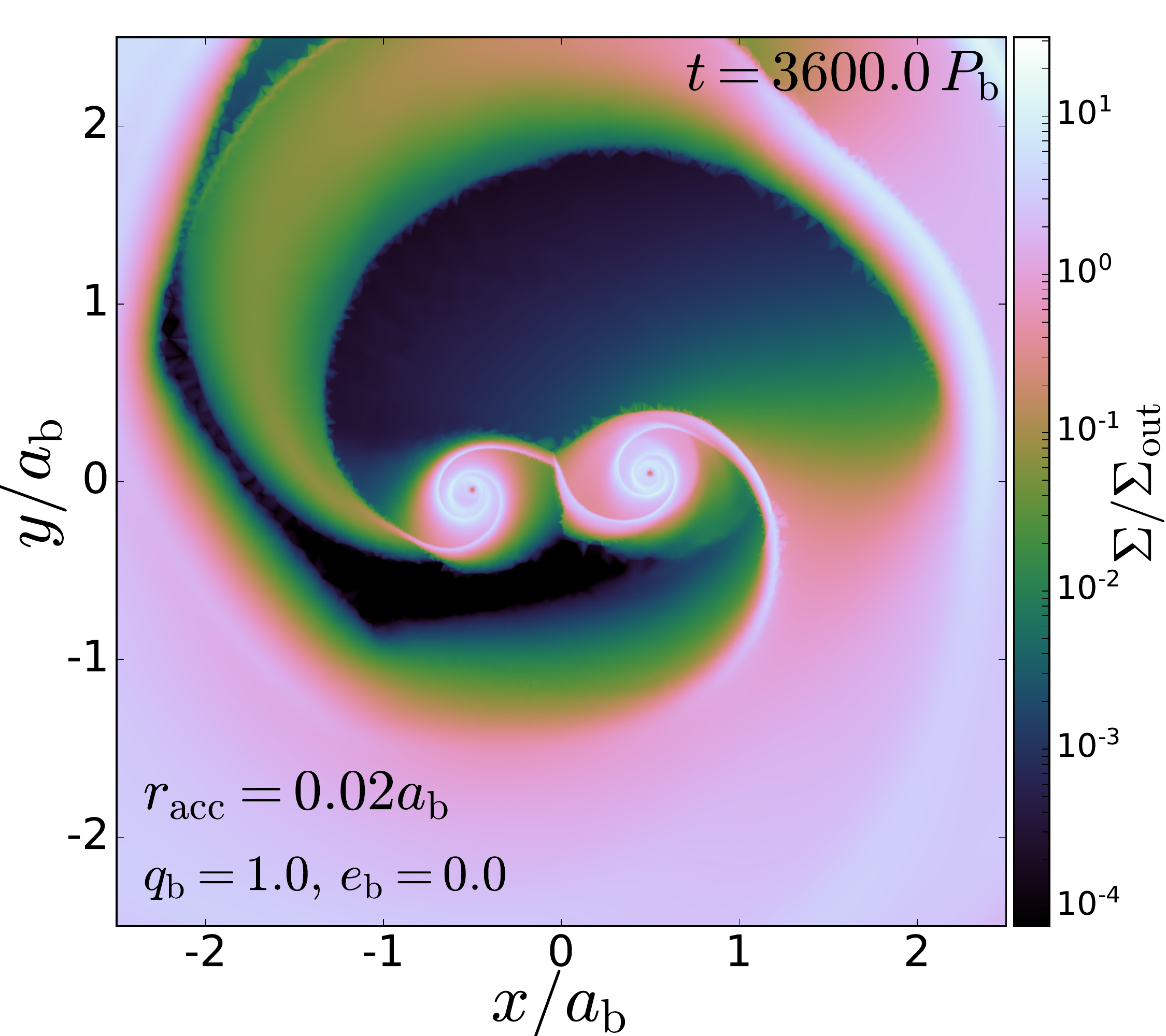}
\includegraphics[width=0.3\textwidth]{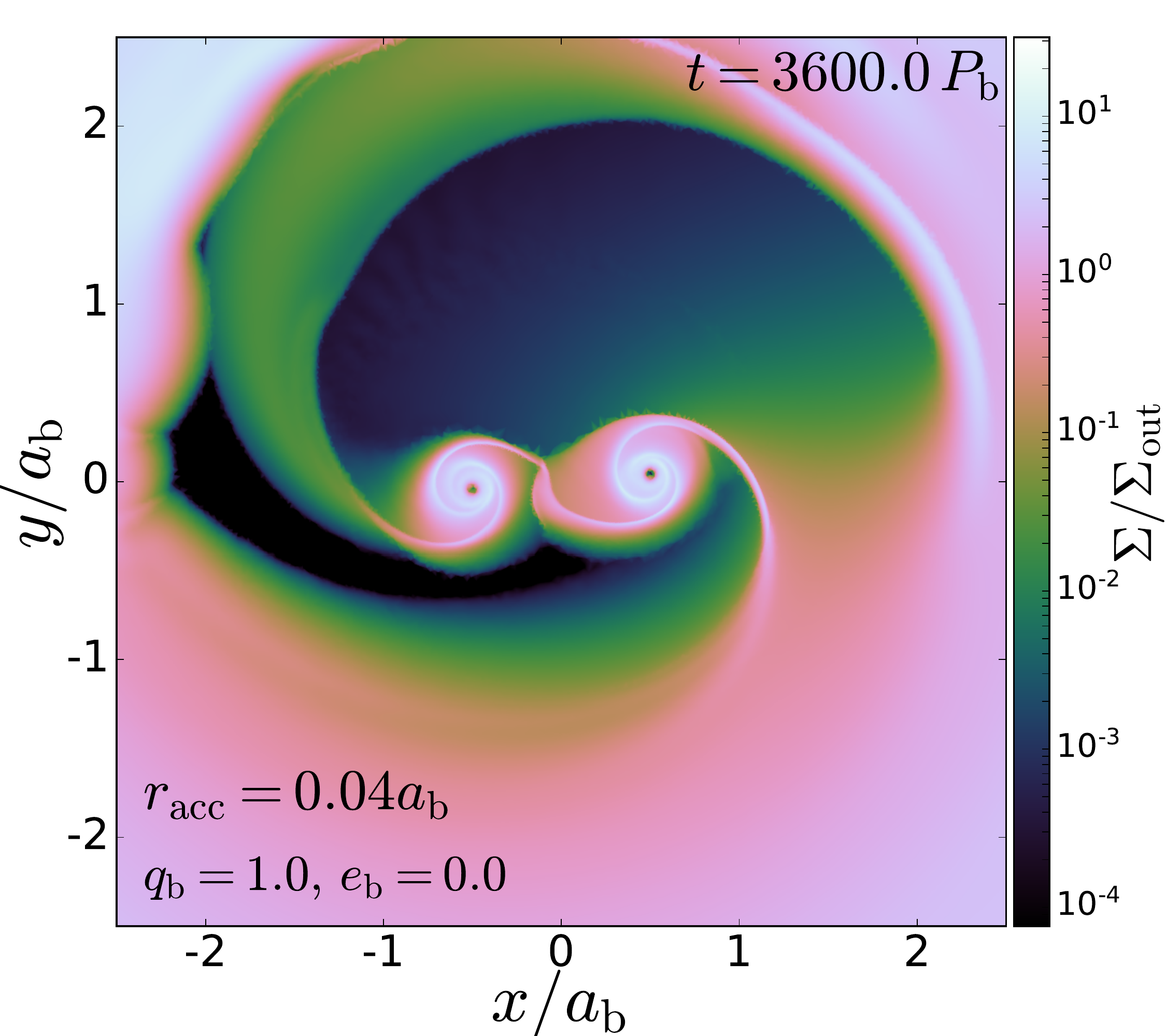}
\includegraphics[width=0.3\textwidth]{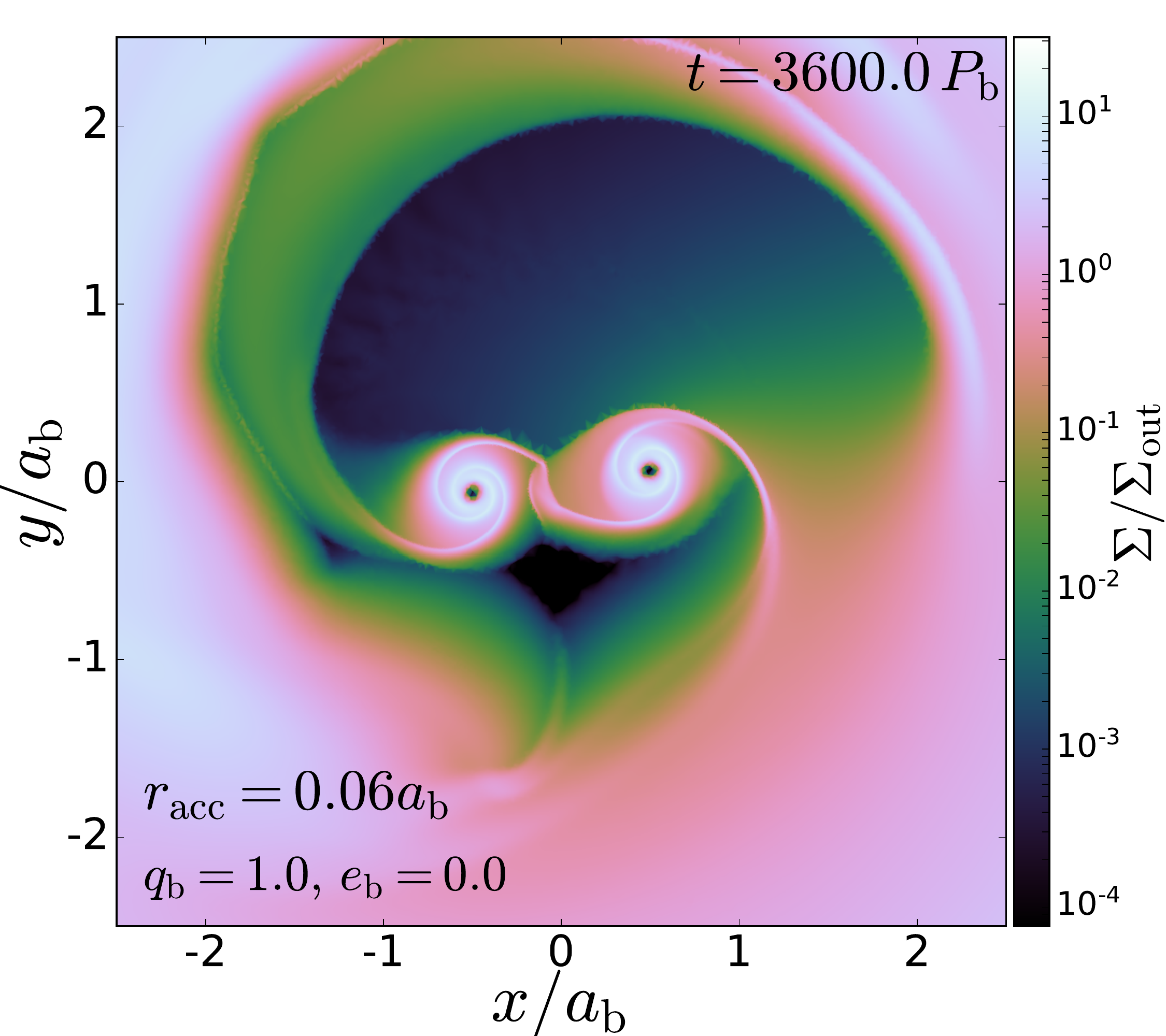}
\vspace{-0.05in}
\centering
\includegraphics[width=0.46\textwidth]{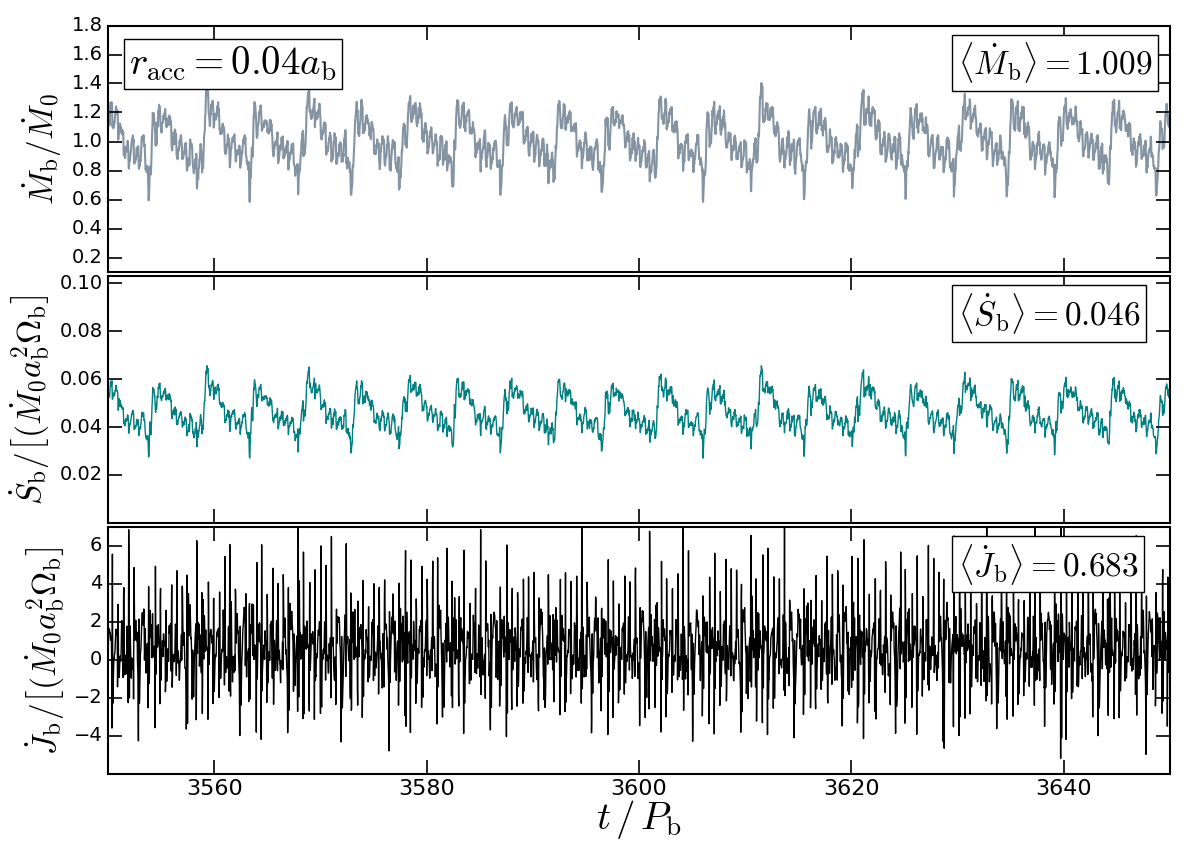}
\includegraphics[width=0.46\textwidth]{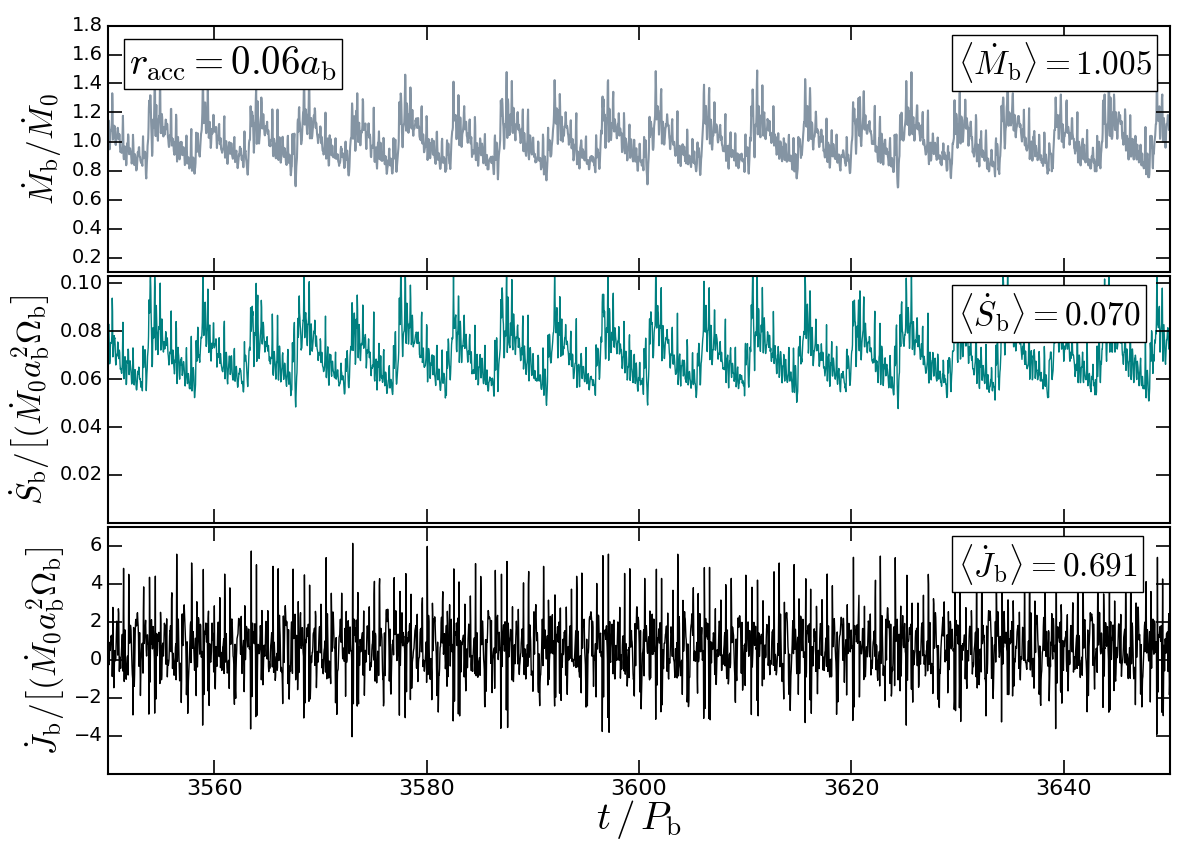}
\vspace{-0.05in}
\caption{Mass and angular momentum transfer for a circular binary for different values of the accretion (sink) radius $r_{\rm acc}$.
The top panels depict the gas surface density rendered at $t=3600\,P_{\rm b}$ for $r_{\rm acc}=0.02a_{\rm b}$ (left),
$r_{\rm acc}=0.04a_{\rm b}$ (middle) and $r_{\rm acc}=0.06a_{\rm b}$ (right).
The bottom panels show the  joint evolution of $\dott{M}_{\rm b}$, $\dot{S}_{\rm b}$ and $\dott{J}_{\rm b}$  over the interval $[3550\,P_{\rm b},3650\,P_{\rm b}]$
for $r_{\rm acc}=0.04a_{\rm b}$ (middle) and $r_{\rm acc}=0.06a_{\rm b}$ (right).
Over the same time interval, all simulations exhibit highly consistent behavior in $\dott{M}_{\rm b}$ and 
$\dott{J}_{\rm b}$, with $\langle\dott{J}_{\rm b}\rangle/\langle\dott{M}_{\rm b}\rangle\approx0.67-0.68$ (see Figure~\ref{fig:angmom_components}). The net
spin torque $\langle\dott{S}_{\rm b}\rangle$ increases almost linearly with $r_{\rm acc}$, while $\langle\dott{J}_{\rm b}\rangle$ is not affected in a significant way.
\label{fig:convergence_sinkradius}}
\vspace{-0.05in}
\end{figure*}
%%%%%%%%%%%%%%%%%%%%%%%%%%%%%%%%%%%%%%%%%%%%%%%%%%%
%%%%%%%%%%%%%%%%%%%%%%%%%%%%%%%%%%%%%%%%%%%%%%%%%%%

Finally, we assume that 
radial transport is dominated by viscosity, and use  
\begin{equation}\label{eq:radial_velocity1}
\begin{split}
\vel_{R}^{\rm (visc)}(R)&=\frac{1}{R\Sigma}\frac{\partial}{\partial R}\left(\nu\Sigma R^3\frac{d\Omega}{dR}\right)
\left[\frac{d}{dR}(R^2\Omega)\right]^{-1}\\
&=-3 \alpha h_0^2\Omega_{\rm b}a_{\rm b} \left(\frac{R}{a_{\rm b}}\right)^{-1/2}\left[\frac{R}{\Sigma}\frac{d\Sigma}{dR}+1\right]
\end{split}
\end{equation}
as an estimate of the initial radial velocity, where we have taken $\Omega\approx\sqrt{\mathcal{G}M_{\rm b}/R}=\Omega_{\rm b}(R/a_{\rm b})^{-3/2}$.
For our initial density profile (Equation~\ref{eq:density_profile}), Equation~(\ref{eq:radial_velocity1}) roughly matches the radial velocity profile of a steady-supply disk, 
\begin{equation}\label{eq:radial_velocity2}
\vel_{R}^{\rm (steady)}=-\frac{\dot{M}_0}{2\pi R\Sigma}~~,
\end{equation}
for $R\gtrsim15a_{\rm b}$. We have also experimented with setting $\vel_{R}=\vel_{R}^{\rm (steady)}$ as an initial condition and have found that
a quasi-steady state can be reached faster. We
stress that any axisymmetric initial condition is an inadequate representation of  the gas dynamics for $R\lesssim8a_{\rm b}$, and 
that the exact choice of the initial condition in this region is not critical to our results. It is of importance, however, that the simulation be evolved for a viscous time 
at a radius where the CBD can be considered axisymmetric and treated approximately as a conventional $\alpha$-disk. For 
an integration time of $t_{\rm int}=3000P_{\rm b}$, the disk is viscously relaxed out to a radius 
$R_{\rm rel}=a_{\rm b}[(9/2)\pi\alpha h_0^2(t_{\rm int}/P_{\rm b})]^{2/3}\approx 12 a_{\rm b}$ (\citealp{raf16}; ML16).
Therefore, outside this radius, the surface density will not have time to relax and will partially retain its initial configuration. It is then
desirable that, already at $t=0$, the accretion rate profile roughly satisfies $\dott{M}(R)\approx \dott{M}_0$ for $R\gtrsim20a_{\rm b}$; the initial
condition (Equation~\ref{eq:density_profile}) meets this requirement.

\subsubsection{Steady-state tests}\label{sec:steady_state}
In order to test how close our initial condition is to a true steady state,
we carry out three test integrations with $e_{\rm b} =0$: ($i$) we run an {\footnotesize AREPO} simulation with an open, diode-like
boundary condition at $R=a_{\rm b}$ for $3000P_{\rm b}$ and then ``release'' the boundary
to allow for direct accretion onto the binary, integrating for additional 500 orbits (see ML16); 
($ii$) we allow for direct accretion onto the ``stars'' immediately at the start-up
of the simulation and integrate for 500 orbits;
($iii$) we use the results of the {\footnotesize PLUTO} simulation from MML17 at $t=3000P_{\rm b}$, remapping it onto an {\footnotesize AREPO}
 grid and rescaling the units such that measured accretion rate at $R=15 a_{\rm b}$ equals unity; then we proceed to evolve this setup integrating,
as in cases ($i$) and ($ii$), for another 500 orbits. The accretion rates corresponding to these three tests are depicted in 
Figure~\ref{fig:accretion_rates}.
In all  cases, after gas crosses the $R=a_{\rm b}$ boundary, only 5-20 orbits are needed for the CSDs to form. Another 20-30 orbits
are needed for the binary to start accreting with an averaged rate equal to $\dott{M}_0$ to within a few percent.  Eventually,  $\dott{M}_0$
is matched to within $1\%$ in all cases, although this can take from 100 orbits (top panel) to 300 orbits (bottom panel). Overall, these
three simulations are indistinguishable from each other, and we conclude that once quasi-steady-state is reached, the initial
conditions become irrelevant.

From here on, we define  {\it quasi-steady-state} as the state of the simulation
for which the averaged accretion rate onto the binary $\langle \dott{M}_{\rm b}\rangle$
over an interval $T=100P_{\rm b}$ is equal to the mass supply rate $\dott{M}_0$ to within 2$\%$ and where
the time averaged mass accretion profile in the CBD $\langle\dott{M}_{\rm d}\rangle(R)$
is also $\approx \dott{M}_0$ (i.e, flat to within an RMS tolerance of $5\%$).
Unless stated otherwise, all simulations presented henceforth include a preliminary phase ($3000P_{\rm b}$) in
which the disk is evolved using a diode boundary condition at $R_{\rm in}=a_{\rm b}(1+e_{\rm b})$.
If a flat $\langle\dott{M}_{\rm d}\rangle$ profile  is reached within that time period,
 the boundary is removed and the simulation is resumed in
the self-consistent fashion described in the preceding paragraphs.

%%%%%%%%%%%%%%%%%%%%%%%%%%%%%%%%%%%%%%%%
\section{SIMULATION RESULTS}\label{sec:results}
%
%%%%%%%%%%%%%%%%%%%%%%%%%%%%%%%%%%%%%%%%%%%%%%%
\subsection{Results for a Circular Binary}\label{sec:circular_binary}
Here, we explore the accretion behavior of a circular, equal-mass binary. 
We focus on the simulation output after 3500 binary orbits.

\subsubsection{Angular momentum transfer}%%%%%%%%%%%%%%%%%%%%%%%%%%%%%
The simulation output is processed to produce time series for $\dott{M}_{\rm b}$ and $\dot{q}_{\rm b}$,
 the specific torques $dl_{\rm b}/dt|_{\rm grav}$ and $dl_{\rm b}/dt|_{\rm acc}$  and the spin change rate $\dot{S}_{\rm b}\equiv\dott{S}_1+\dott{S}_2$
  (see Section~\ref{sec:direct_torques}).  These quantities and the total $\dott{J}_{\rm b}$ are shown in Figure~\ref{fig:angmom_components}.
for a time interval of $200P_{\rm b}$. We find that in quasi-steady-state, the time averaged $\dott{J}_{\rm b}$ is given by 
\begin{equation}
\langle\dott{J}_{\rm b}\rangle  \approx 0.676 \dott{M}_0\Omega_{\rm b}a_{\rm b}^2~~.
\end{equation}
Of the five different contributions to $\dott{J}_{\rm b}$,
two quantities dominate over the rest: the mass change $\langle\dott{M}_{\rm b}\rangle\approx \dott{M}_0$ (first panel)
and  specific gravitational torque  
$\langle \dot{l}_{\rm b, grav}\rangle \approx 1.567(\dott{M}_0/{M}_{\rm b})\Omega_{\rm b}a_{\rm b}^2$  (third panel). 

The stationary nature of the $\dott{M}_{\rm b}$ and $\dott{J}_{\rm b}$ means that, in an averaged sense, a constant amount of angular momentum
is transferred to the binary by the disk per unit accreted mass. This constant is the eigenvalue of the accretion flow \citep[e.g.][]{pac91,pop91}, i.e.,
\begin{equation}
\frac{\langle\dott{J}_{\rm b}\rangle}{\langle\dott{M}_{\rm b}\rangle}\equiv l_0~~.
\end{equation}
Our result in Figure~\ref{fig:angmom_components} implies
\begin{equation}\label{eq:accretion_eigenvalue}
l_0 \vphantom{\frac{1}{2}}\big|_{e_{\rm b}=0} \approx\lcirc\Omega_{\rm b}a_{\rm b}^2~~.
\end{equation}
The eigenvalue $l_0$ can also be obtained by computing $\langle\dott{J}_{\rm d}\rangle(R)$, the angular momentum current as a function of radius in the
CBD (Equation~\ref{eq:angular_momentum_balance}).
We compute $\dott{J}_{\rm d}(R)$ from the combination of 
$\dott{J}_{\rm d,adv}$, $\dott{J}_{\rm d,visc}$ and  $\dott{J}_{\rm d,grav}$, in turn computed from the time averaged angular momentum
flux maps $F_J$ as described in Section~\ref{sec:angmom_transfer}. We compute these angular momentum flux maps
from simulation snapshots produced at a high output rate (15 snapshots per orbit) followed by time-averaging. Figure~\ref{fig:angmom_currents}
shows the result. Importantly, we see that 
the net angular momentum current $\langle\dott{J}_{\rm d}\rangle(R)$ (dark red)
is nearly flat for $R$ between $a_{\rm b}$ and $10a_{\rm b}$. 
Furthermore, $\langle\dott{J}_{\rm d}\rangle(R)$ is very close to 
$\langle\dott{J}_{\rm b}\rangle$. 
In other words, all of the angular momentum crossing the CBD is received by the binary. This provides
convincing evidence that the exchange of mass and
angular momentum between the CBD and the binary are in true quasi-steady-state, and that our computed eigenvalue (Equation~\ref{eq:accretion_eigenvalue}) is reliable.

%%%%%%%%%%%%%%%%%%%
\paragraph{Dependence on sink radius}
In our simulations, the "stars" behave as sink particles. In most of our simulations, we adopt the 
``stellar radius'' of  $r_{\rm acc}=0.02 a_{\rm b}$, but it is natural to evaluate how sensitive our results are
to the choice of $r_{\rm acc}$. Recently, \citet{tan17} posited that the sign of the net angular momentum transfer rate
$\langle \dott{J}_{\rm b}\rangle$ (note that \citealp{tan17} only computed  $\langle \dott{L}_{\rm b}\rangle$ and did not include spin torques)
can depend critically on the way gas is drained from the computational domain by the stars (see Section~\ref{sec:previous_work} below).
We test the influence of the accretion algorithm on our results by carrying out two additional simulations with $r_{\rm acc}=0.04a_{\rm b}$ 
and $r_{\rm acc}=0.06a_{\rm b}$ (the softening lengths are updated accordingly, but all other parameters remain unchanged).
The results of this test are depicted in Figure~\ref{fig:convergence_sinkradius}, where we contrast the gas density field (to illustrate
the effect of a larger sink radius) with the evolution of $\dott{M}_{\rm b}$, $\dott{S}_{\rm b}$ and $\dott{J}_{\rm b}$. The mean values
$\langle\dott{J}_{\rm b}\rangle$ and $\langle\dott{M}_{\rm b}\rangle$ are robust against $r_{\rm acc}$, but $\langle\dott{S}_{\rm b}\rangle$ grows almost proportionally to  $r_{\rm acc}$. 
Since $\langle\dott{J}_{\rm b}\rangle$ appears unmodified, the increase in $\langle\dott{S}_{\rm b}\rangle$ must be
accompanied by a decrease in the torque $\mu_{\rm b}\langle\dot{l}_{\rm b, grav}\rangle$ (not shown).
  {Indeed, over the interval $[3500,3700]P_{\rm b}$, the specific gravitational torque 
$\langle\dot{l}_{\rm b, grav}\rangle$ is $1.534(\dott{M}_0/M_{\rm b})\Omega_{\rm b}a_{\rm b}^2$ when $r_{\rm acc}=0.04a_{\rm b}$
and $1.482(\dott{M}_0/M_{\rm b})\Omega_{\rm b}a_{\rm b}^2$ when $r_{\rm acc}=0.06a_{\rm b}$, both smaller than the value
of $1.567(\dott{M}_0/M_{\rm b})\Omega_{\rm b}a_{\rm b}^2$ in our fiducial simulation (Fig.~\ref{fig:angmom_components}).
The anisotropic accretion torque $\mu_{\rm b}\langle\dot{l}_{\rm b, acc}\rangle$ remains negligible in all cases}. For all the values of $r_{\rm acc}$ explored,
$\langle\dott{S}_{\rm b}\rangle$ is a small contribution to the total transfer of angular momentum to the binary $\langle\dott{J}_{\rm b}\rangle$.  
Presumably, a much larger
value of $r_{\rm acc}$ than explored here -- one that forbids the formation of CSDs -- could make $\langle\dott{S}_{\rm b}\rangle$ an important contributor to 
$\langle\dott{J}_{\rm b}\rangle$, one that competes with  $\mu_{\rm b}\dot{l}_{\rm b, grav}$, perhaps to the point of erasing the positive torque contribution
of the CSDs, turning all the positive gain in orbital angular momentum into a growth in spin.

%%%%%%%%%%%%%%%%%%%
\subsubsection{Understanding the Origin of the Positive Torque on the Binary: Spatial Distribution of Gravitational Torques}
A major result of this paper (see Figs.~\ref{fig:angmom_components}-\ref{fig:angmom_currents}),
is that $\langle \dot{J}_{\rm b}\rangle\simeq \langle \dot{J}_{\rm d}\rangle>0$, i.e., an accreting binary {\it gains} angular
momentum from a steady-supply disk. 
From Figure~\ref{fig:angmom_components}, we see that a major contribution to $\langle \dot{J}_{\rm b}\rangle$
comes from the total gravitational torque $\mu_{\rm b} \langle \dot{l}_{\rm b,grav}\rangle$ (third panel), which is positive,
 in apparent contradiction with what one might expect from theoretical work \citep{gol80,art94,lub96}, since 
 the linear theory of Lindblad torques of \citet{gol79} predicts that an outer disk
exerts negative torques on the inner binary.
Note that our result for the angular momentum current $\dott{J}_{\rm d,grav}$ in the CBD 
(Figure~\ref{fig:angmom_currents}, yellow curve) is indeed consistent with this theoretical expectation. 
However, a self-consistent calculation of $\langle \dot{J}_{\rm b}\rangle$ needs to include the entire distribution of gas, and not only that of the CBD.

%%%%%%%%%%%%%%%%%%%%%%%%%%%%%%%%%%%%%%%%%%%%%%%%%%%
%%%%%%%%%%%%%%%%%%%%%%%%%%%%%%%%%%%%%%%%%%%%%%%%%%%
\begin{figure*}[h!]
\vspace{-0.1in}
\centering
\includegraphics[width=0.47\textwidth]{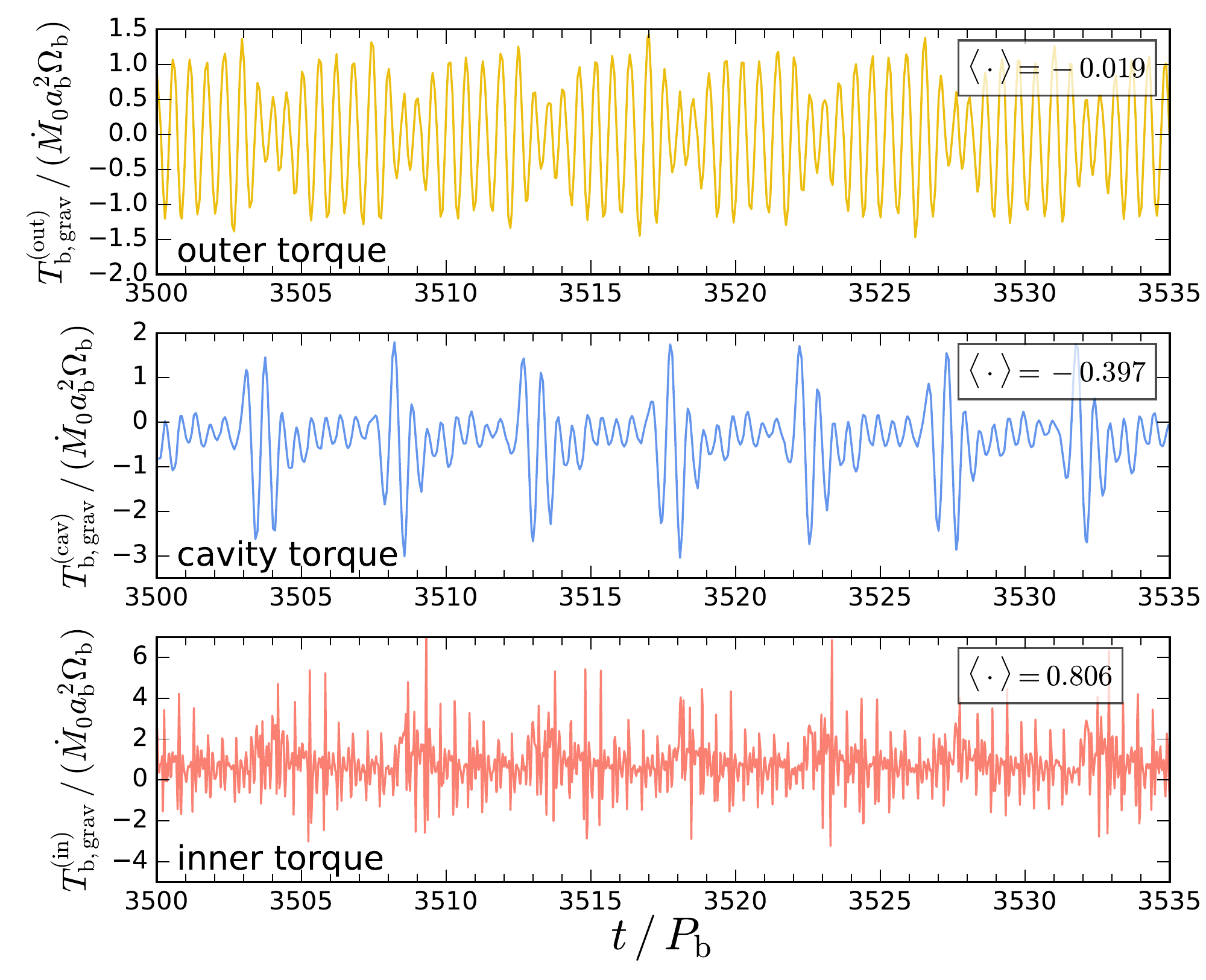}
\includegraphics[width=0.47\textwidth]{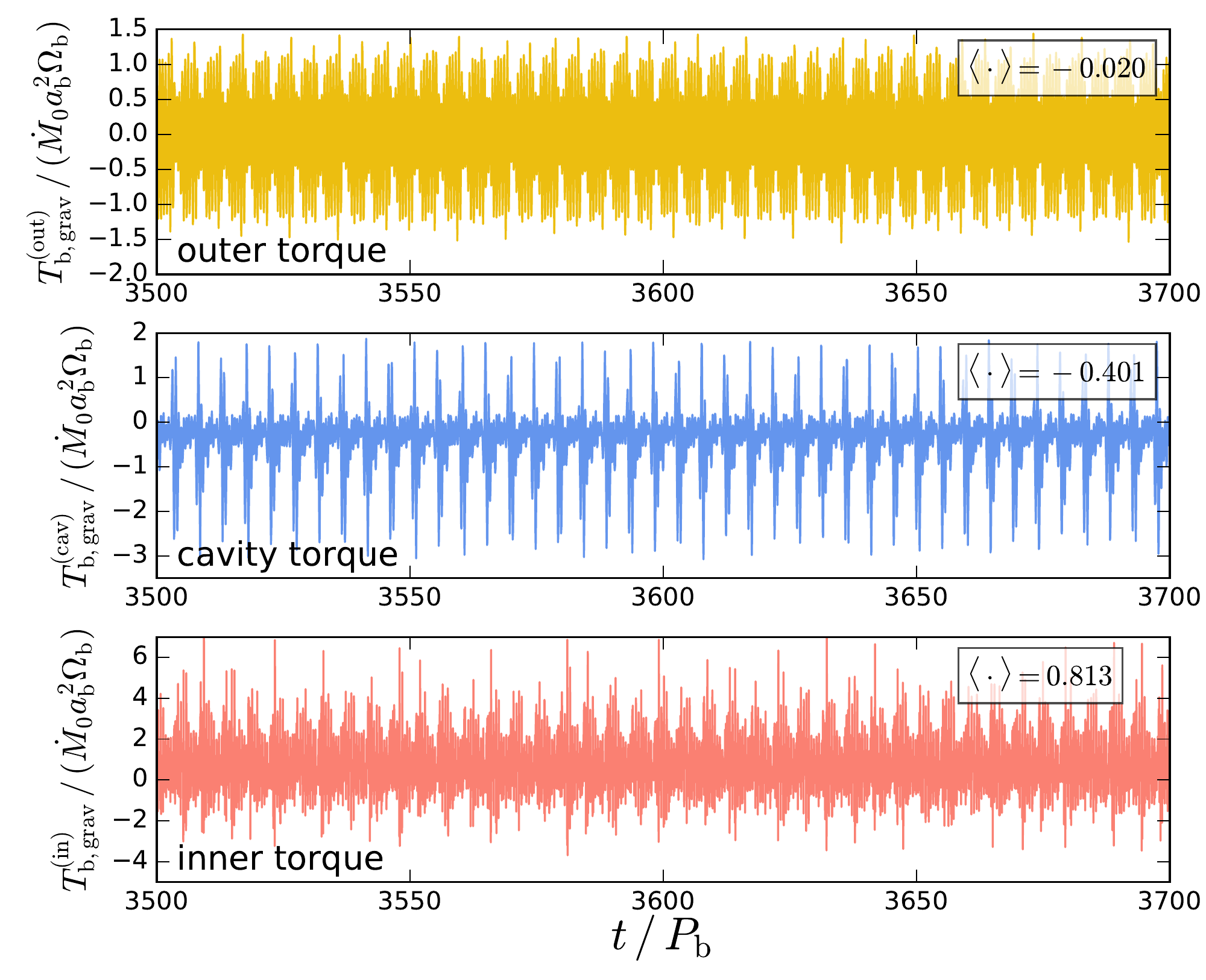}
\vspace{-0.07in}
\caption{Gravitational torques dissected according to distance $R$ from the binary barycenter (Section~\ref{sec:initial_setup}) as a function of time,
covering the time interval between 3500 and 3535 $P_{\rm b}$  (left panels) and
between 3500 to 3700 $P_{\rm b}$  (right panels). The outer torque (yellow, top panels) corresponds to
the torque exerted by gas cells located at $R>R_{\rm cav})$ (see Equation~\ref{eq:cavity_size}); 
the cavity torque (blue, middle panels) corresponds to the total torque exerted by cells with radii  $a_{\rm b}<R<R_{\rm cav}$;
the inner torque (red, bottom panels) corresponds to cells with  $R<a_{\rm b}$. The mean values of these torques are indicated in each panel.
Both the cavity and outer torques are negative, collectively amounting to $\approx -0.42\dott{M}_0\Omega_{\rm b}a_{\rm b}^2$,
in agreement with $\dott{J}_{\rm d,grav}$ at $R=a_{\rm b}$ (Figure~\ref{fig:angmom_currents}). 
A net negative gravitational torque due to gas {\it outside} $R=a_{\rm b}$ is in agreement with
theoretical expectations and past numerical work. The net positive sign of $\langle T_{\rm b,grav}\rangle$ is due to the large inner torque $\langle T_{\rm b,grav}^{\rm (in)}\rangle$.
 {Over 200 orbits (right panels), the net gravitational torque $\langle T_{\rm b,grav}\rangle=0.392\dott{M}_0\Omega_{\rm b}a_{\rm b}^2$ is equal to 
the specific torque $\langle \dot{l}_{\rm b,grav}\rangle=1.567(\dott{M}_0/M_{\rm b})\Omega_{\rm b}a_{\rm b}^2$  
(third panel of Fig.~\ref{fig:angmom_components}) multiplied by the reduced mass $\mu_{\rm b}=M_{\rm b}/4$.}
The inner torque exhibits rapid variability on time-scales $\sim P_{\rm b}/100$ and its mean value, when properly measured, can double in magnitude the negative torque due to the outer/cavity gas.
Comparison between the left and right panels shows that a few tens of orbits are  sufficient to capture the basic stationary behavior of these torque time series.
\vspace{-0.03in}
\label{fig:spatial_torques_dissected}}
\end{figure*}
%%%%%%%%%%%%%%%%%%%%%%%%%%%%%%%%%%%%%%%%%%%%%%%%%%%
%%%%%%%%%%%%%%%%%%%%%%%%%%%%%%%%%%%%%%%%%%%%%%%%%%%
%%%%%%%%%%%%%%%%%%%%%%%%%%%%%%%%%%%%%%%%%%%%%%%%%%%
%%%%%%%%%%%%%%%%%%%%%%%%%%%%%%%%%%%%%%%%%%%%%%%%%%%
\begin{figure*}[h!]
\vspace{-0.15in}
\centering
\includegraphics[width=0.87\textwidth]{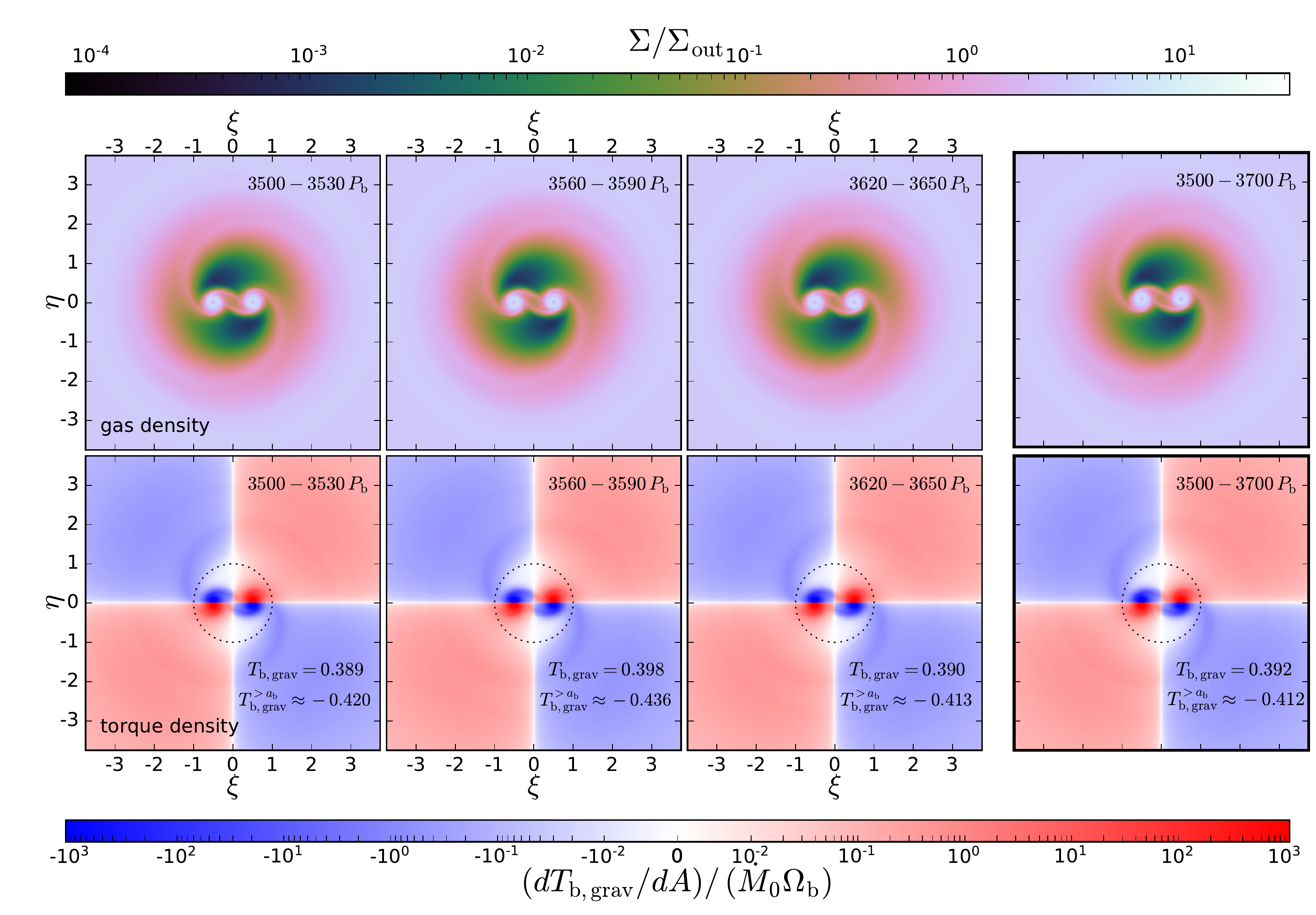}
\vspace{-0.04in}
\caption{Stacked (averaged) maps  of the 
mass density $\Sigma$ (top panels) and the gravitational torque density  $dT_{\rm b, grav}/d{\rm A}$  (bottom panels)
for accretion onto a circular binary.
From left to right, the stacked maps for three different time
windows of length $30 P_{\rm b}$ (corresponding to 450 snapshots) are shown:
3500 to 3530 orbits, 3560 to 3590 orbits and 3620 to 3650 orbits.
The differences between these maps are virtually unnoticeable, suggesting 
that $\simeq30$ orbits is sufficient integration time
to capture the essential behavior of the accreting binary. 
The integrated  torque maps over area outside
$R=a_{\rm b}$ is $T_{\rm b}^{>a_{\rm b}}\simeq-0.41\dott{M}_0\Omega_{\rm b}a_{\rm b}^2$, 
which is roughly consistent over time.
The last column shows the stacking of 3000 reconstructed maps over 200 binary orbits; again, the overall morphology
 appears to be unchanged over longer timescales.
\label{fig:spatial_torques_averaged}}
\end{figure*}
%%%%%%%%%%%%%%%%%%%%%%%%%%%%%%%%%%%%%%%%%%%%%%%%%%%
%%%%%%%%%%%%%%%%%%%%%%%%%%%%%%%%%%%%%%%%%%%%%%%%%%%

Figure~\ref{fig:spatial_torques_dissected} depicts the dissection of the total gravitational torque $T_{\rm b, grav}=\mu_{\rm b} \dot{l}_{\rm b,grav}$ 
into 3 components (from top to bottom): (i) the outer disk torque $T_{\rm b,grav}^{\rm (out)}$  ($R>R_{\rm cav}$; with $R_{\rm cav}$ as defined in Equation~\ref{eq:cavity_size}),
(ii) the ``cavity+streams'' torque $T_{\rm b,grav}^{\rm (cav)}$
($a_{\rm b}\leq R\leq R_{\rm cav}$); and (iii) the ``inner''  torque $T_{\rm b,grav}^{\rm (in)}$  ($R<a_{\rm b}$). The short-term
evolution of these torques is depicted on the left panels of the figure, and the long-term evolution is shown on the right.
On average, the sum of these 3 different
torques must equal  $\langle T_{\rm b, grav}\rangle=\mu_{\rm b} \langle\dot{l}_{\rm b,grav}\rangle=0.392 \dott{M}_0a_{\rm b}^2\Omega_{\rm b}$.

From the torque dissection, we confirm that the gravitational torque acting on the binary due to the material outside $R=a_{\rm b}$ is indeed negative, but that the material {\it inside}
$R=a_{\rm b}$ exerts a positive torque that nearly doubles in magnitude the negative outer torque. We also
 find that the oscillation amplitudes of these three torque contributions can be much larger than their mean values. In order to capture both the rapid sign changes in $T_{\rm b,grav}^{\rm (out)}$,
 $T_{\rm b,grav}^{\rm (cav)}$ and $T_{\rm b,grav}^{\rm (in)}$ as well as their long-term averages,  a fine sampling in time as well as a long-term integration are required. We find that a sample rate of 15 
 snapshots per binary orbit captures $T_{\rm b,grav}^{\rm (out)}$ and $T_{\rm b,grav}^{\rm (cav)}$ adequately since they vary on timescales of $\sim P_{\rm b}$; however, since $T_{\rm b,grav}^{\rm (in)}$ can vary on timescales comparable
to the orbital period within the CSDs, we need a sampling rate of $\sim100$ snapshots per orbit to capture such rapid variability.
This rapid variability is difficult to capture from post-processing analysis of simulation snapshots, but is automatically accounted for by the strategy
described in Section~\ref{sec:direct_torques}.

%%%%%%%%%%%%%%%%%%%%%%%%%%%%%%%%%%%%%%%%%%%%%%%
%%%%%%%%%%%%%%%%%%%%%%%%%%%%%%%%%%%%%%%%%%%%%%%
\begin{figure*}[ht!]
\vspace{-0.05in}
\centering
\includegraphics[width=0.83\textwidth]{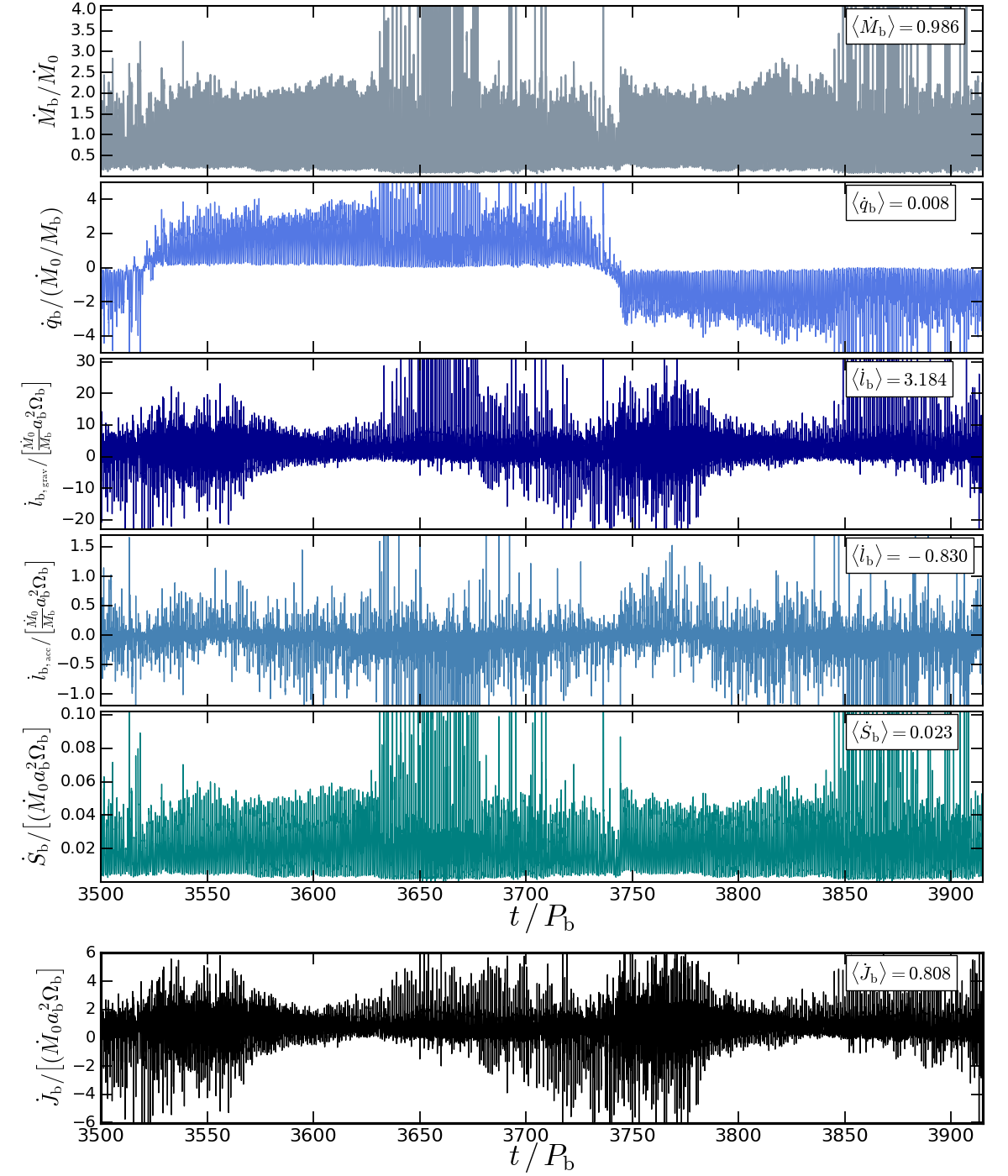}
\caption{The five different contributions to angular momentum transfer rate and its combined effect $\dott{J}_{\rm b}$
 {(Equations~\ref{eq:angmom_change} and~\ref{eq:reduced_mass_change})},
for an eccentric ($e_{\rm b}=0.6$) binary. Panels are the same as in Figure~\ref{fig:angmom_components}, with the only difference being the time axis, which covers 415 binary orbits (as opposed to the 200 of the $e_{\rm b}=0$ case). The extent of the time axis is chosen to roughly match the alternation period in mass accretion 
(see change in the sign of $\dot{q}_{\rm b}$ in second panel from top). The accretion eigenvalue in this case
is $l_0\approx\lecc \Omega_{\rm b}a_{\rm b}^2$.
We have also explored the evolution of these same quantities
for the time interval $[3985\,P_{\rm b},4400\,P_{\rm b}]$, confirming that the $\langle\dott{M}_{\rm b}\rangle$ and $\langle\dott{J}_{\rm b}\rangle$
are unchanged. 
\label{fig:angmom_components_ecc}}
\end{figure*}
%%%%%%%%%%%%%%%%%%%%%%%%%%%%%%%%%%%%%%%%%%%%%%%
%%%%%%%%%%%%%%%%%%%%%%%%%%%%%%%%%%%%%%%%%%%%%%%

The contribution of the cavity region to the gravitational torque can be better understood if we explore the spatial
distribution of the torque in two dimensions. For this, we construct gravitational torque density maps.
Evidently, since $-\mathbf{T}_{\rm b, grav}=\mathbf{T}_{\rm d, grav}$ (the torque exerted on the disk by the binary), the torque per unit
area can be computed directly from the cell-centered values of density and gravitational acceleration:
\begin{equation}\label{eq:torque_density}
\begin{split}
\frac{d\mathbf{T}_{\rm b, grav}}{d{\rm A}} (\mathbf{r})=-\frac{d\mathbf{T}_{\rm d, grav}}{d{\rm A}} (\mathbf{r})&=
\Sigma(\mathbf{r}) \mathbf{r}\times \nabla\Phi_{\rm b}(\mathbf{r})\\
&\approx -\Sigma_k \mathbf{r}_k\times \mathbf{a}_k~~,
\end{split}
\end{equation}
where $\mathbf{a}_k$ is the gravitational acceleration of the gas and the subscript $k$ denotes evaluation at the center of a  cell.

To explore
the long-term effect of the nonaxisymmetric features in the circumbinary gas, we can take time averages of both the density and torque density fields. This is shown in Figure~\ref{fig:spatial_torques_averaged},
where we have stacked the density field $\Sigma$ and the torque per unit  area (Equation~\ref{eq:torque_density}) (15 snapshots per orbit)  rotated by an angle that matches
the rotation of the binary. The coordinates axes in this figure are thus $\xi$ and $\eta$, related to the original coordinates by the area-preserving transformation
$x=a_{\rm b}(\xi \cos f_{\rm b} - \eta\sin f_{\rm b})$ and $y=a_{\rm b}(\xi\sin f_{\rm b} + \eta\cos f_{\rm b})$~\footnote{%%%%%%%%%%%%%%%%%%%%%%
Since the determinant of this transformation's Jacobian matrix is unity, the density and torque density are not altered by the rotation of the coordinate system.}. %%%%%%%%%%%%%%%%%%%%%%%%%%%%%%%%%%%%%%%%%%%%%%%%%%%%%%%%%%%%%%%%%%%%%%%%%%%
We carry out this average over intervals of $\sim30$ orbits in duration, at different times in the simulation. In addition, we perform the average
over 200 orbits (rightmost panels of Figure~\ref{fig:spatial_torques_averaged}). All the averaging sets produce essentially the same result, implying that most of the variability takes place on timescales shorter than 30 orbits.
Figure~\ref{fig:spatial_torques_averaged} reveals that the streamers are the dominant nonaxisymmetric feature that persists in time.
From the torque density maps (bottom panels) it can be seen that different portions of the gas can ``torque up'' (regions in red)
or ``torque down'' (regions in blue) the binary, but 
only the persistent nonaxisymmetric features contribute to the non-zero net torque.

%%%%%%%%%%%%%%%%%%%%%%%%%%%%%%%%%%%%%%
\subsection{Results for an Eccentric Binary}\label{sec:eccentric_binary}
We repeat the numerical simulations of Section~\ref{sec:circular_binary}, this time for  binary eccentricity  $e_{\rm b}=0.6$.
Several differences between these two calculations are expected. First, the lump that modulates the accretion every $\sim5$ binary orbits
when $e_{\rm b}=0$ (e.g., \citealp{mac08,shi12}; ML16; MML17) is no longer present (ML16; for these values of $\alpha$ and $h_0$, MML17
find that the lump disappears at $e_{\rm b}\gtrsim0.05$). Second, accretion will be preferential and alternating (\citealp{dun15}; ML16), i.e., the primary
and secondary will alternate over precessional timescales which one receives most of the mass. Third, the gas distribution and kinematics in the region $R\lesssim a_{\rm b}$
differ significantly from the circular case (material enters and leaves this region in a periodic fashion, 
see Fig. 2 of ML16), affecting the amount of 
torque that this region can provide to counterbalance the torque from the CBD. 

\subsubsection{Angular momentum transfer}%%%%%%%%%%%%%%%%%%%%%%%%%%%%%
In Figure~\ref{fig:angmom_components_ecc} we show the total angular momentum transfer rate 
$\dott{J}_{\rm b}$ and its various contributions for a $q_{\rm b}=1$, $e_{\rm b}=0.6$ binary.
As expected, we find that $\dott{M}_{\rm b}$ is no longer modulated at  a frequency
of $\sim\tfrac{1}{5}\Omega_{\rm b}$, but at the dominant frequency $\sim \Omega_{\rm b}$
(ML16). There are variations over much longer timescales: the mass ratio change $\dot{q}_{\rm b}$ flips
 signs every $\sim415\,P_{\rm b}$. The gravitational specific torque $\dot{l}_{\rm b, grav}$ is again positive, and larger than in the circular case. 
The contribution of the anisotropic accretion torque $\mu_{\rm b}\langle \dot{l}_{\rm b,acc}\rangle \approx -0.21\dott{M}_0\Omega_{\rm b}a_{\rm b}^2$ 
 is no longer negligible. As in the circular case, the contribution of the spin torque $\langle \dott{S}_{\rm b}\rangle$
   to the total torque $\langle \dott{J}_{\rm b}\rangle$ is small ($\sim2\%$).

%%%%%%%%%%%%%%%%%%%%%%%%%%%%%%%%%%%%%%%%%%%%%%%%%%%
%%%%%%%%%%%%%%%%%%%%%%%%%%%%%%%%%%%%%%%%%%%%%%%%%%%
\begin{figure}[h!]
\centering
\includegraphics[width=0.45\textwidth]{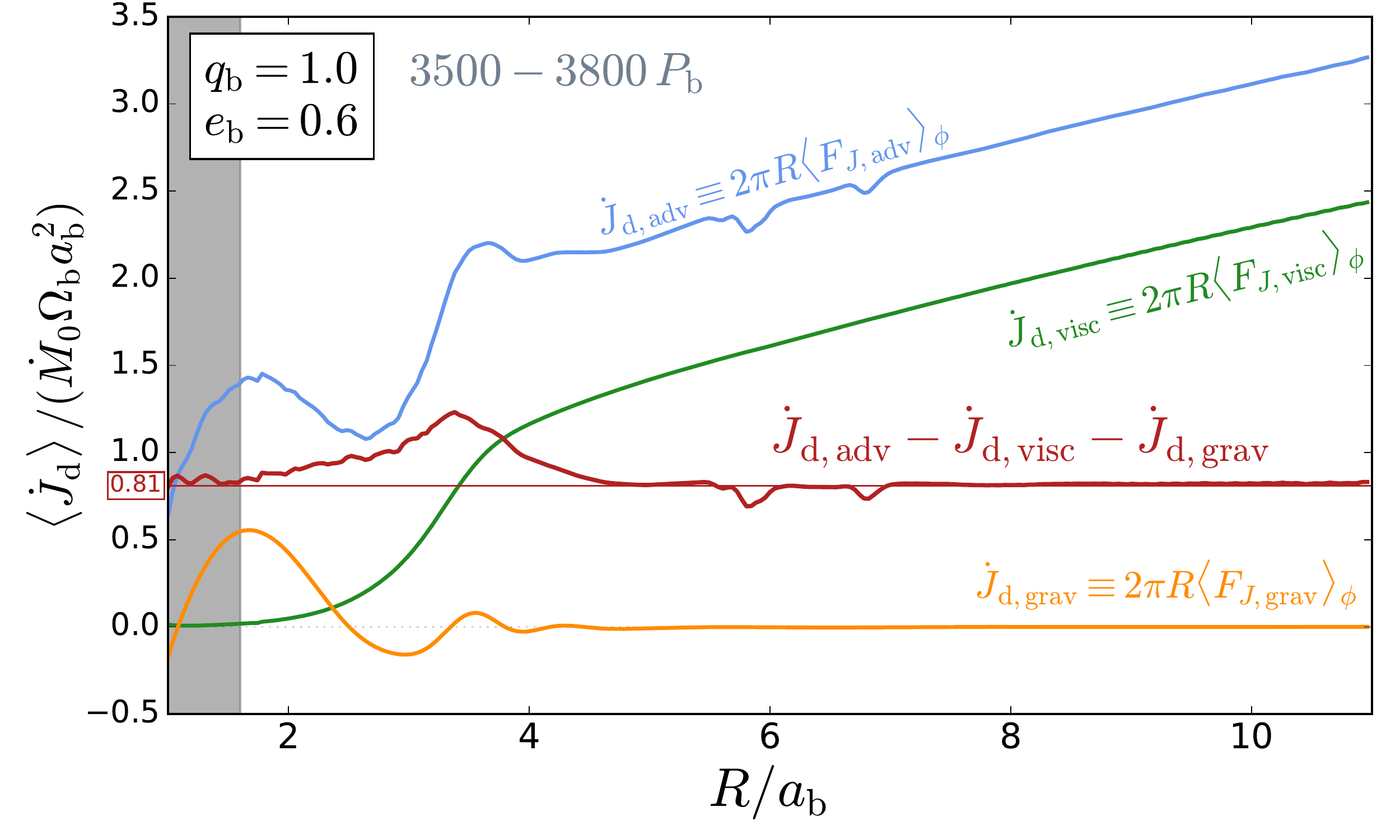}
\caption{Similar to Figure~\ref{fig:angmom_currents}, except for an eccentric binary ($e_{\rm b}=0.6$).
 { The thin red
line indicates the eigenvalue $l_0=\lecc a_{\rm b}^2\Omega_{\rm b}$} (Equation~\ref{eq:eigenvalue_ecc}) obtained from the evolution of the accreting
eccentric binary  (Figure~\ref{fig:angmom_components_ecc}). The gray region represents the portion of the computational domain
left out by the diode-like boundary at $R=a_{\rm b}(1+e_{\rm b})$ in the {\footnotesize PLUTO} simulations of MML17 and is depicted here only for reference.
The $30\%$ discrepancy at $R\approx3a_{\rm b}$ is likely due to a combination of insufficiently dense sampling in time and
the incomplete coverage of apsidal precession period of the disk, in addition to the intrinsic difficulties of estimating the advective flux
of angular momentum from reconstructed primitive variables.
\label{fig:angmom_currents_ecc}}
\end{figure}
%%%%%%%%%%%%%%%%%%%%%%%%%%%%%%%%%%%%%%%%%%%%%%%%%%%
%%%%%%%%%%%%%%%%%%%%%%%%%%%%%%%%%%%%%%%%%%%%%%%%%%%

 %%%%%%%%%%%%%%%%%%%%%%%%%%%%%%%%%%%%%%%%%%%%%%%%%%%
%%%%%%%%%%%%%%%%%%%%%%%%%%%%%%%%%%%%%%%%%%%%%%%%%%%
\begin{figure*}[h!]
\centering
\includegraphics[width=0.49\textwidth]{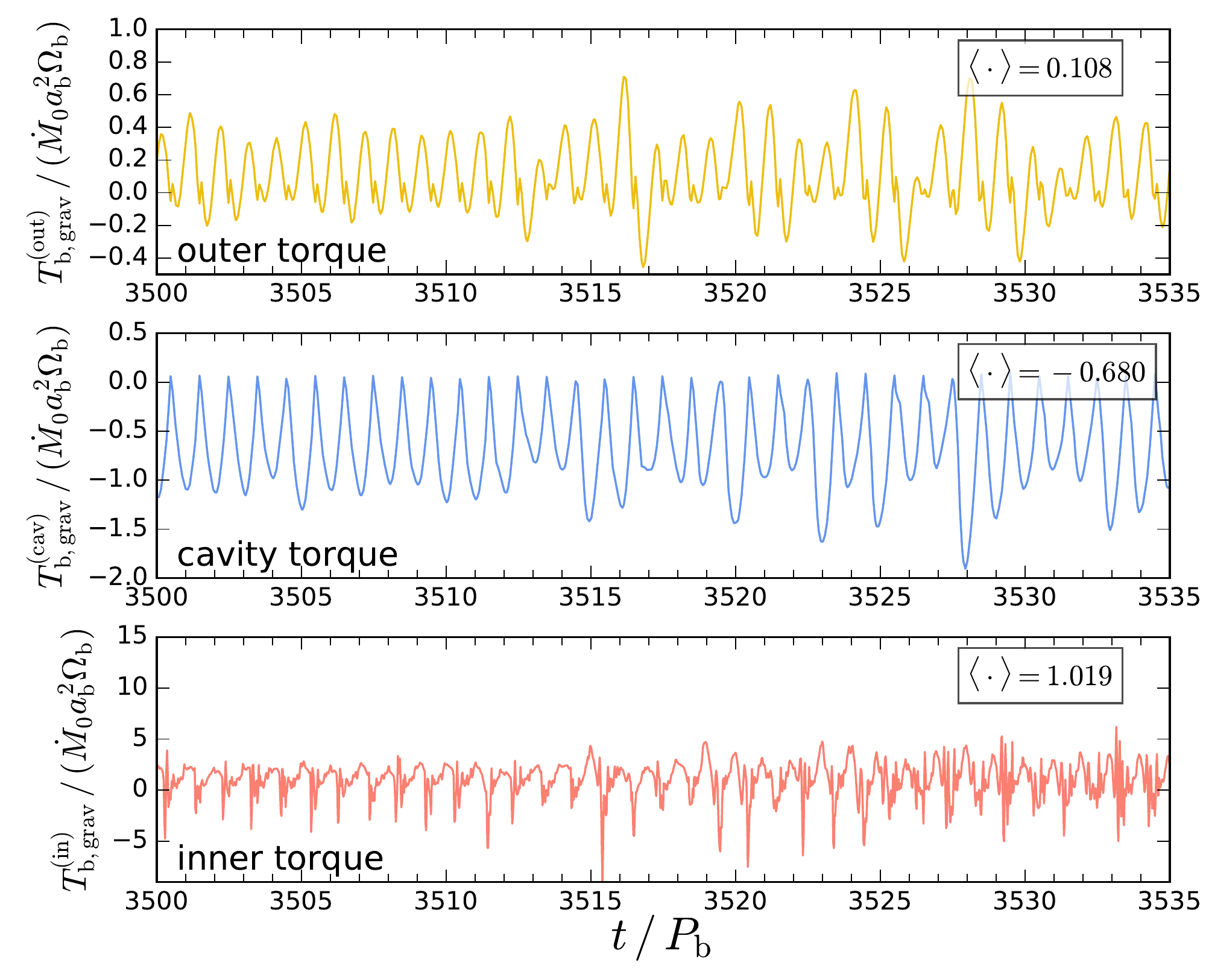}
\includegraphics[width=0.49\textwidth]{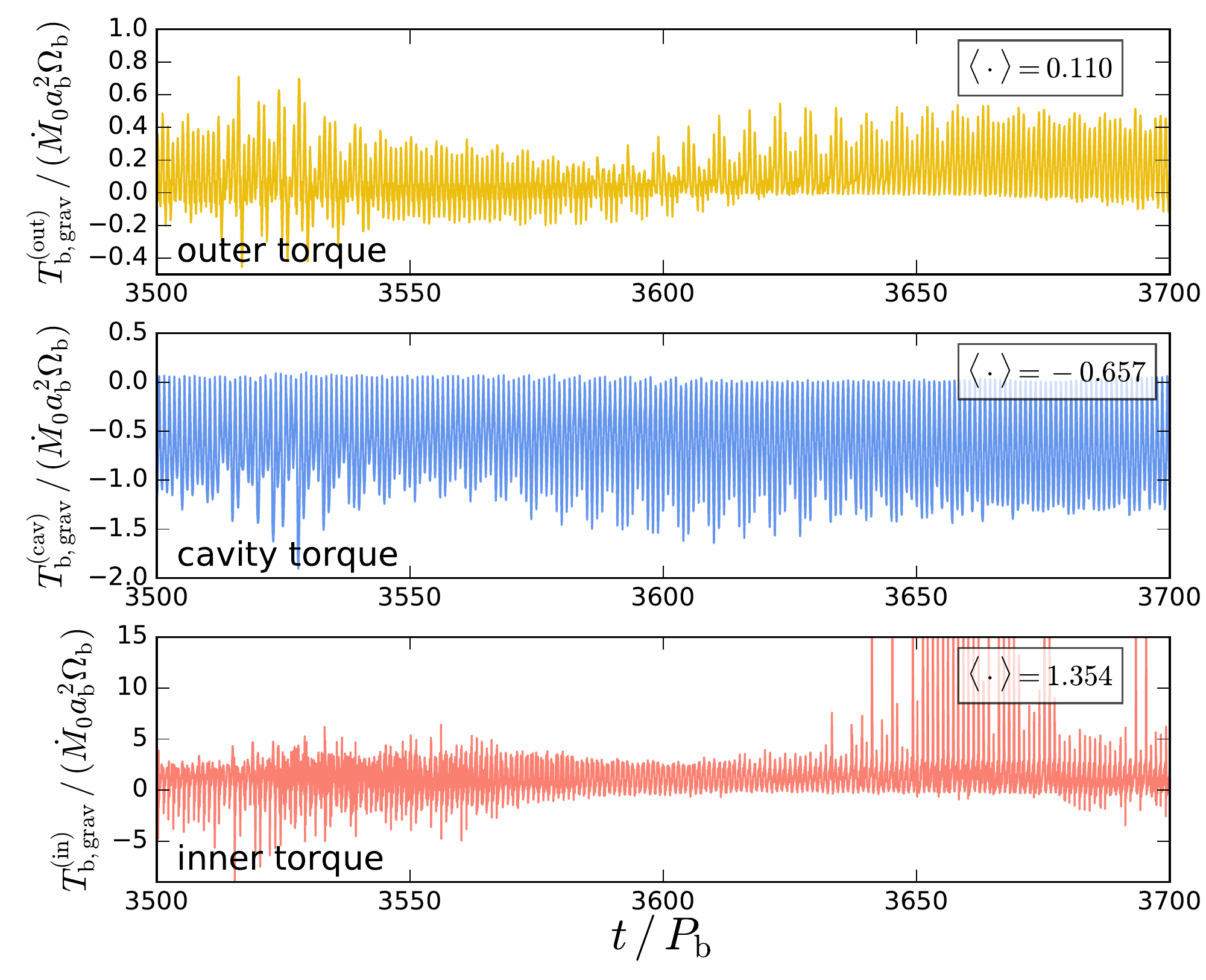}
\caption{Similar to Figure~\ref{fig:spatial_torques_dissected},  this time for an eccentric binary ($e_{\rm b}=0.6$).
The total gravitational torque is dissected into 
 an outer component (yellow, top panels) for $R>R_{\rm cav}$, 
a cavity components (blue, middle panel)  for $a_{\rm b}(1+e_{\rm b})<R\leq R_{\rm cav}$,
  and  an inner component  (red, bottom panels)  for $R\leq a_{\rm b}(1+e_{\rm b})$. 
 The gravitational torque due to gas outside the boundary $R=a_{\rm b}(1+e_{\rm b})$ is negative, but smaller in magnitude than
 the positive torque due to gas inside that boundary. 
  {Over 200 orbits (right panels), the net gravitational torque $\langle T_{\rm b,grav}\rangle=0.807\dott{M}_0\Omega_{\rm b}a_{\rm b}^2$ is equal to 
the specific torque $\langle \dot{l}_{\rm b,grav}\rangle=3.229(\dott{M}_0/M_{\rm b})\Omega_{\rm b}a_{\rm b}^2$  
(extracting the interval $[3500,3700]P_{\rm b}$ from the third panel of Fig.~\ref{fig:angmom_components_ecc}) and multiplied by the reduced mass $\mu_{\rm b}=M_{\rm b}/4$.}
\label{fig:spatial_torques_dissected_ecc}}
\end{figure*}
%%%%%%%%%%%%%%%%%%%%%%%%%%%%%%%%%%%%%%%%%%%%%%%%%%%
%%%%%%%%%%%%%%%%%%%%%%%%%%%%%%%%%%%%%%%%%%%%%%%%%%%
%%%%%%%%%%%%%%%%%%%%%%%%%%%%%%%%%%%%%%%%%%%%%%%%%%%
%%%%%%%%%%%%%%%%%%%%%%%%%%%%%%%%%%%%%%%%%%%%%%%%%%%
\begin{figure*}[h!]
\centering
\includegraphics[width=0.87\textwidth]{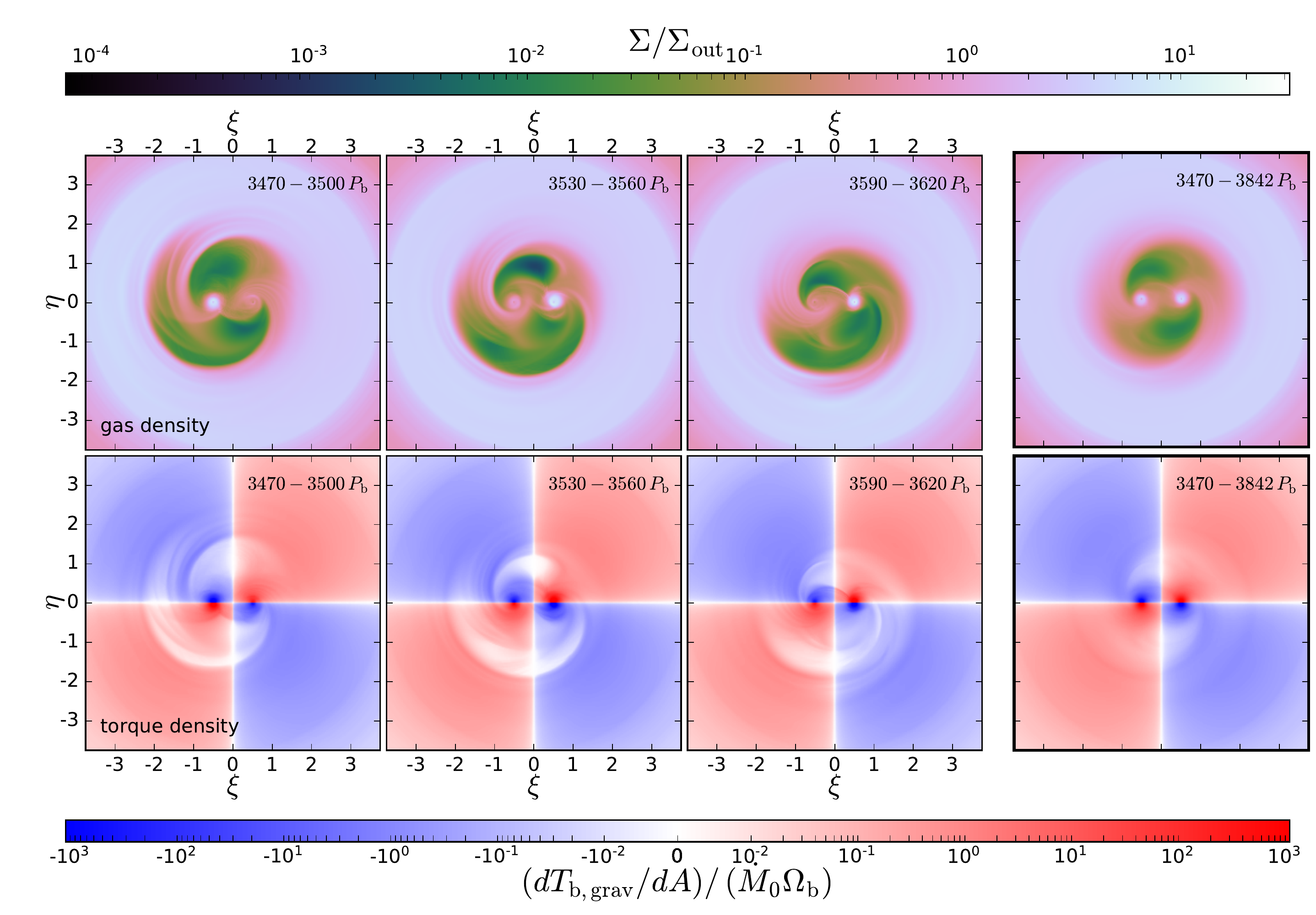}
\caption{Stacking of mass density and torque density maps, similar to Figure~\ref{fig:spatial_torques_averaged},
this time for a binary with $e_{\rm b}=0.6$. Stacking is carried out in the rotating-pulsating coordinate
frame of the eccentric binary (see text); in this frame, the primary and secondary remain along the horizontal axis, with fixed coordinates
$\xi=-0.5$ and $\xi=0.5$ respectively.
From left to right, we show three time windows of length $30\,P_{\rm b}$:  3470 to 3500 orbits,  3530 to 3560 orbits, and
3500 to 3620 orbits. In contrast to Figure~\ref{fig:spatial_torques_averaged}, these stacked frames are evidently not exhibiting stationarity. Most notable
is the asymmetry in the surface density of the CSDs: in the first map, the disk around the primary is mass loaded,
while the one around the secondary is barely visible; in the second map,  a transition has taken place, where the secondary
disk contains more mass; in the third map, the secondary disk is loaded, while the cicum-primary disk is empty.
This transition is consistent with what is observed in the time series of $\dott{M}_1$ and $\dott{M}_2$ (ML16) and in the long-term
behavior of $\dot{q}_{\rm b}$ in Figure~\ref{fig:angmom_components_ecc}. This long-term variability is apparent in the change of integrated torque outside the binary $T_{\rm b, grav}^{>1.6a_{\rm b}}$.
 The last column on the right depicts the stacking of 4935 maps of
$\Sigma$ and  $dT_{\rm b, grav}/dA$ over 372 binary orbits; most of the variability is removed, and the surface density
map shows a great degree of symmetry, with both CSDs containing similar amounts of mass. 
\label{fig:spatial_torques_averaged_ecc}}
\vspace{0.08in}
\end{figure*}
%%%%%%%%%%%%%%%%%%%%%%%%%%%%%%%%%%%%%%%%%%%%%%%%%%%
%%%%%%%%%%%%%%%%%%%%%%%%%%%%%%%%%%%%%%%%%%%%%%%%%%%

%%%%%%%%%%%%%%%%%%%%%%%%%%%%%%%%%%%%%%%%%%%%%%%%%%%
\begin{figure*}[th!]
\vspace{-0.1in}
\centering
\includegraphics[width=0.78\textwidth]{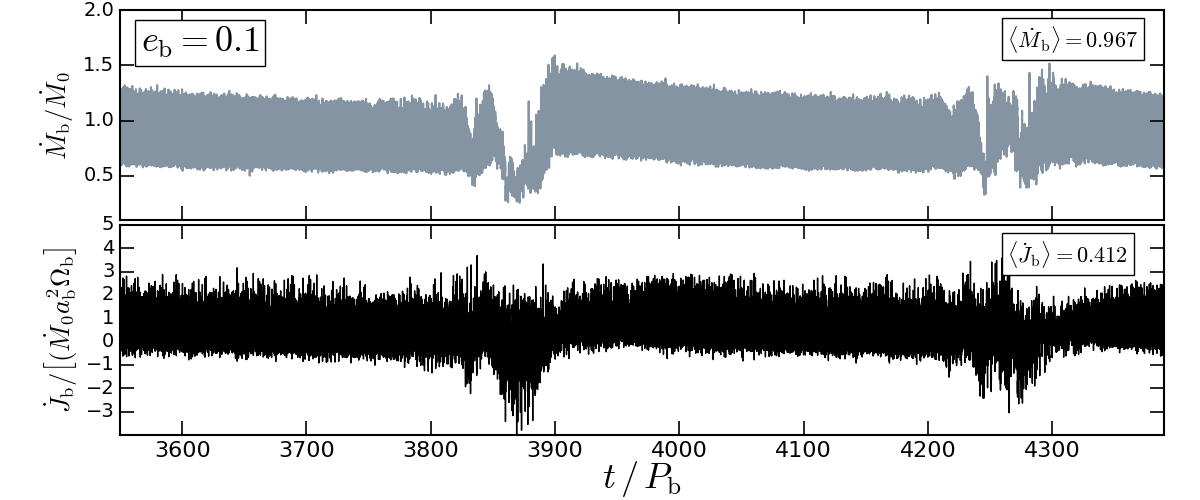}
\includegraphics[width=0.78\textwidth]{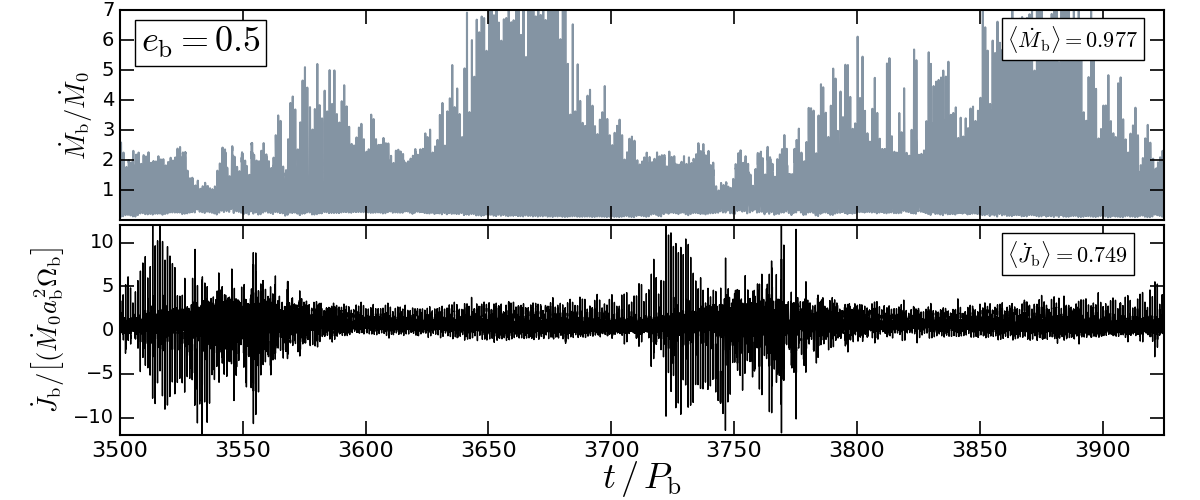}
\caption{Similar to Figs.~\ref{fig:angmom_components} and~\ref{fig:angmom_components_ecc}, 
but for binaries with $e_{\rm b}=0.1$ and $e_{\rm b}=0.5$, and only showing 
$\dott{M}_{\rm b}$ (gray) and $\dott{J}_{\rm b}$  (black). The accretion eigenvalue is $l_0\approx0.43\Omega_{\rm b}a_{\rm b}^2$
for $e_{\rm b}=0.1$ and  $l_0\approx0.78\Omega_{\rm b}a_{\rm b}^2$ for $e_{\rm b}=0.5$.
\label{fig:angmom_components_examples}}
\vspace{-0.1in}
\end{figure*}
%%%%%%%%%%%%%%%%%%%%%%%%%%%%%%%%%%%%%%%%%%%%%%%%%%%%%%%%%%%%%%%%%%
%%%%%%%%%%%%%%%%%%%%%%%%%%%%%%%%%%%%%%%%%%%%%%%%%%%%%%%%%%%%%%%%%%

As in Section~\ref{sec:circular_binary}, we can compute the accretion eigenvalue $l_0$ 
directly from the result shown in Figure~\ref{fig:angmom_components_ecc}, giving
\begin{equation}\label{eq:eigenvalue_ecc}
l_0 \vphantom{\frac{1}{2}}\big|_{e_{\rm b}=0.6} \approx\lecc\Omega_{\rm b}a_{\rm b}^2~~.
\end{equation}
We also compute the net angular momentum current in the CBD 
$\dott{J}_{\rm d}(R)$ (Section~\ref{sec:angmom_transfer})  averaged
over 300 orbits, as shown in Figure~\ref{fig:angmom_currents_ecc}. 
We see that $\langle\dott{J}_{\rm d}\rangle(R)$ is approximately constant (independent of $R$) and equal to 
$\langle \dott{J}_{\rm b}\rangle$ , indicating that the global angular momentum balance is achieved. 
In this case, the match between $\langle\dott{J}_{\rm d}\rangle/\dott{M}_0$ and
$l_0=0.81\Omega_{\rm b}a_{\rm b}^2$ is remarkable for $R>4a_{\rm b}$ and  $R<2a_{\rm b}$.

%%%%%%%%%%%%%%%%%%%
\subsubsection{Spatial Distribution of Gravitational Torques}
As in the case of the circular binary, the eccentric binary gains angular momentum, which, as before, is largely a result of
the positive net gravitational torque
 $\langle T_{\rm b,grav}\rangle=\mu_{\rm b}\langle \dot{l}_{\rm b, grav}\rangle \approx 0.796\dott{M}_0\Omega_{\rm b}a_{\rm b}^2$. 
As before, we study the spatial distribution of gravitational torques in the simulation by splitting
$T_{\rm b,grav}$ into three components: $T_{\rm b,grav}^{\rm (out)}$, 
$T_{\rm b,grav}^{\rm (cav)}$ and $T_{\rm b,grav}^{\rm (in)}$ (Figure~\ref{fig:spatial_torques_dissected_ecc}).
We see a similar behavior to that of a circular binary (Figure~\ref{fig:spatial_torques_dissected}): the
sum $T_{\rm b,grav}^{\rm (out)}+T_{\rm b,grav}^{\rm (cav)}$ is  negative (dominated by $T_{\rm b,grav}^{\rm (cav)}$)
while the positive sign of the total torque $\langle T_{\rm b,grav}\rangle$ is entirely due to the positive sign of $T_{\rm b,grav}^{\rm (in)}$. The  torque time series
shown in Figure~\ref{fig:spatial_torques_dissected_ecc} exhibit some degree of regularity, but 
they differ from the nearly exact periodicity seen in Figure~\ref{fig:spatial_torques_dissected}. As noted above, the accretion flow onto eccentric binaries is modulated over long secular timescales, 
 and it is not surprising that a few tens of orbits cannot capture the entire range of time variability.

The fact that tens of orbits is too short a time span to reveal the underlying stationary behavior the accretion flow can be illustrated
by repeating the map-stacking analysis  for the $e_{\rm b}=0.6$ case (see Figure~\ref{fig:spatial_torques_averaged_ecc}; cf. Figure~\ref{fig:spatial_torques_averaged}).
The stacking of these images is trickier than in the circular case:  in order for the two ``stars'' to line-up at all snapshots, we need to 
introduce a rotating-pulsating coordinate system, a transformation that is well known in the study of
 the elliptical restricted three-body problem  \citep[ER3BP; e.g.,][]{nec26,kop63,sze64,sze67,mus14}. The scaled, rotated coordinates $\xi$ and $\eta$ are related
to the barycentric coordinates $x$ and $y$ by: $x=r_{\rm b}(\xi \cos f_{\rm b} - \eta\sin f_{\rm b})$ and
$y=r_{\rm b}(\xi\sin f_{\rm b} + \eta\cos f_{\rm b})$, where $r_{\rm b}=a_{\rm b}(1-e^2_{\rm b})/(1+e_{\rm b}\cos f_{\rm b})$
and $f_{\rm b}$ is the true anomaly of the binary. This transformation is not area preserving, thus scalars must be transformed
by multiplying by the Jacobian of the transformation ($=r_{\rm b}^2$).  In 
Figure~\ref{fig:spatial_torques_averaged_ecc} (top row, left to right), we show $\Sigma$ stacked over the time intervals
$[3470\,P_{\rm b},3500\,P_{\rm b}]$, $[3530\,P_{\rm b},3560\,P_{\rm b}]$, $[3590\,P_{\rm b},3620\,P_{\rm b}]$ and $[3470\,P_{\rm b},3799\,P_{\rm b}]$.
These reveal a striking, inherent
property of accretion onto eccentric binaries: the ``loading" of the CSDs alternates in time, i.e., the gas surface density is higher in one CSD than in the other. 
This disk loading can be directly tied to the sign of $\dot{q}_{\rm b}$ in Fig~\ref{fig:angmom_components_ecc}. 
When we average over 370$P_{\rm b}$ -- almost the entire precessional period identified from the time series of $\dot{q}_{\rm b}$ -- we see that most of the asymmetries disappear 
(the last column of Figure~\ref{fig:spatial_torques_averaged_ecc}), and that the distribution of gas starts to resemble that of the circular case (Figure~\ref{fig:spatial_torques_averaged}). Note
how the shape of the cavity also changes as preferential accretion switches recipients, as it is expected if alternation is caused by the precession of the cavity edge.

We use this rotating system to study the spatial distribution of torques, albeit only in a qualitative manner (Figure~\ref{fig:spatial_torques_averaged_ecc}). Some similarities
can be found with the analogous plot for the circular case (Figure~\ref{fig:spatial_torques_averaged}, bottom row), although the spatial distribution
of torques in the eccentric case is more diffuse and variable. A direct comparison to the spatial torque dissection of Figure~\ref{fig:spatial_torques_dissected_ecc} is not possible,
since the barycentric radial coordinate $R$ changes from snapshot to snapshot in the rotating-pulsating coordinate system.

 %%%%%%%%%%%%%%%%%%%%%%%%%%%%%%%%%%%%%%%%%a
\begin{figure*}[t!]
\centering
\includegraphics[width=0.49\textwidth]{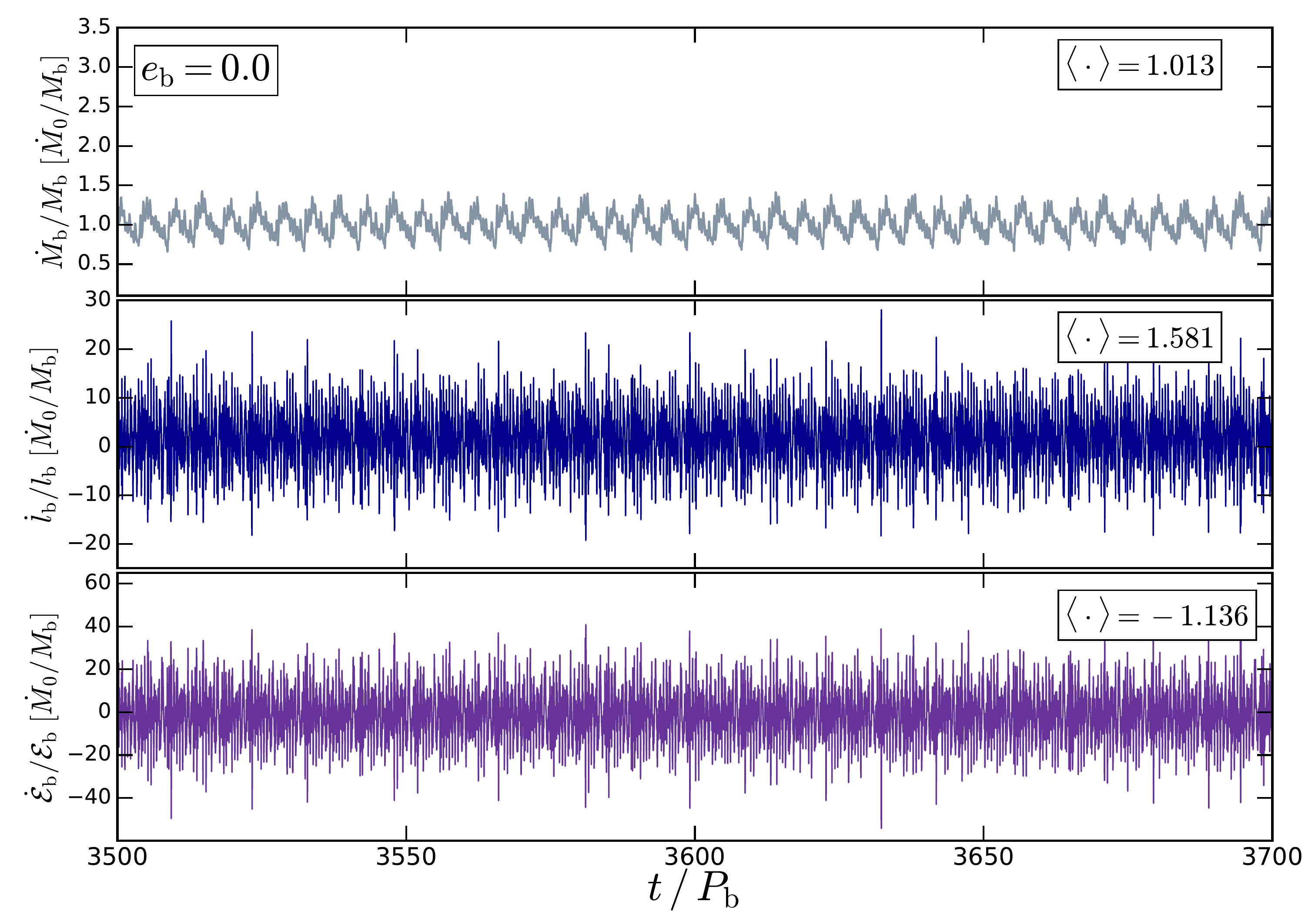}
\includegraphics[width=0.49\textwidth]{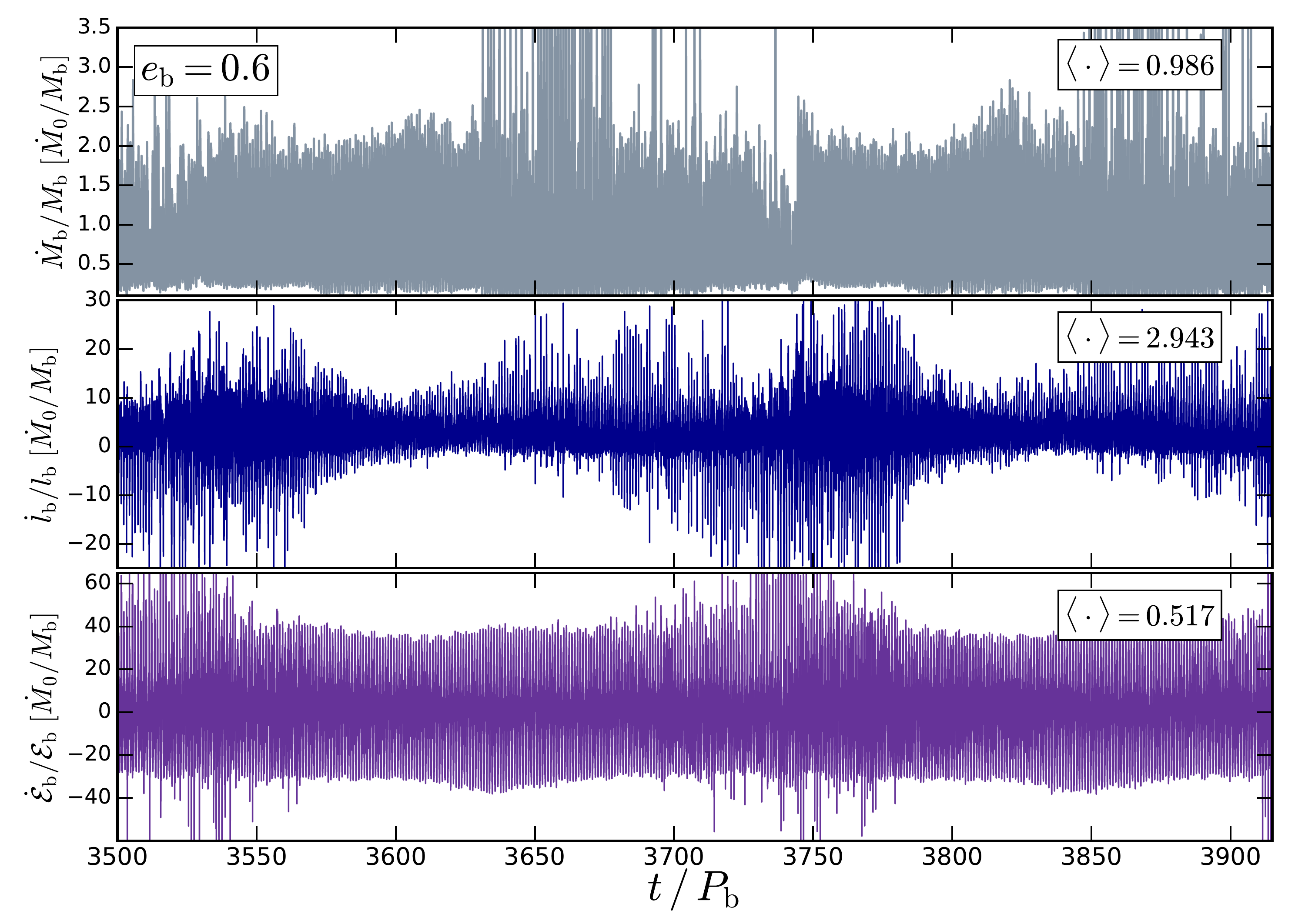}
\caption{Evolution of the rate of change in mass $\dott{M}_{\rm b}/{M}_{\rm b}$ (top), specific angular momentum 
  $\dot{l}_{\rm b}/{l}_{\rm b}$ (middle) and specific energy  $\dott{\mathcal{E}}_{\rm b}/\mathcal{E}_{\rm b}$ (bottom)
  for the $e_{\rm b}=0$ binary (left panels) and the $e_{\rm b}=0.6$ binary (right panels). These quantities determine
  the rate of change of the orbital elements  $\dot{e}_{\rm b}$ and $\dot{a}_{\rm b}$ via
 Eqs.~(\ref{eq:edot}) and~(\ref{eq:adot}).
\label{fig:angmom_energy_change}}
\end{figure*}
%%%%%%%%%%%%%%%%%%%%%%%%%%%%%%%%%%%%%%%%%a

%%%%%%%%%%%%%%%%%%%%%%%%%%%%%%%%%%%%%%%%%%
\subsection{Other Values of Binary Eccentricity}
We have repeated the experiments of Sections~\ref{sec:circular_binary} and~\ref{sec:eccentric_binary}
for other values of the binary eccentricity.
Figure~\ref{fig:angmom_components_examples} shows the results for a low eccentricity binary ($e_{\rm b}=0.1$, top)
and an intermediate eccentricity binary  ($e_{\rm b}=0.5$, bottom). In both cases, the circumbinary cavity lacks the accretion
lump (MML17), and accretion exhibits long-term
alternation. As in Figure~\ref{fig:angmom_components_ecc}, the time interval in the $x$-axis is chosen to cover and entire precessional
period, estimated as the time it takes for $\dot{q}_{\rm b}$ to change sign twice. This period is measured to be $\sim800P_{\rm b}$
for $e_{\rm b}=0.1$ and $\sim400P_{\rm b}$ for $e_{\rm b}=0.5$.
Note that this alternation period does not exactly match the apsidal precession period 
of test particles at the location of the cavity edge  \citep[e.g.,][]{thu17}, 
 since the latter scales with cavity size and eccentricity as
  $(R_{\rm cav}/a_{\rm b})^{7/2}(1+\tfrac{3}{2}e_{\rm b}^2)^{-1}$ (Eq. 7 in ML16), which, if we take Equation~(\ref{eq:cavity_size}) at face value, implies that the precession period scales as $e_{\rm b}^{7/2}$. However,
the precession rate of the inner CBD can be coherent out to $R\sim20a_{\rm b}$ (e.g., Fig. 9 in MML17), and then it does not need to be set by the precession rate at $R_{\rm cav}$,
but rather by some mean rate over that range of radii.

%%%%%%%%%%%%%%%%%%%%%%%%%%%%%%%%%%%%%%%%%%%%%%%%%%%%%%%%%%a
\capstartfalse 
\begin{deluxetable}{ccccccc}
\tabletypesize{\footnotesize}
\tablewidth{0.49\textwidth}
\tablecaption{Summary of simulation results\label{tab:results}}
\tablehead{
\colhead{$e_{\rm b}$} & \colhead{$r_{\rm acc}$} &\colhead{$\langle \dott{J}_{\rm b}\rangle/\langle \dott{M}_{\rm b}\rangle$}&
\colhead{$\langle \dott{S}_{\rm b}\rangle/\langle \dott{M}_{\rm b}\rangle$} & \colhead{$\langle \dot{a}_{\rm b}\rangle$}  
& \colhead{$\langle \dot{e}_{\rm b}\rangle$} & $T$ \\
& \colhead{$[a_{\rm b}]$} &\colhead{$[\Omega_{\rm b}a_{\rm b}^2]$} &\colhead{$[\Omega_{\rm b}a_{\rm b}^2]$} 
& \colhead{$[a_{\rm b}\dott{M}_0/M_{\rm b}]$} & \colhead{$[\dott{M}_0/M_{\rm b}]$} & \colhead{$[P_{\rm b}]$}
}
\startdata
0 & 0.02 & \lcirc & 0.028 & 2.15 & -0.004 & 200 \\
 & 0.04  &0.68 & 0.046 &  2.05 & -0.008 & 200 \\
 &  0.06 & 0.69 & 0.07  & 2.00  & -0.025 & 200  \\
\cline{2-7}\\
0.1 & 0.02 & 0.43 & 0.023 & 0.75 & 2.42 & 800 \\ 
0.5 & 0.02 & 0.78 & 0.025 & 0.95 & -0.20 & 415 \\ 
0.6 & 0.02 & \lecc & 0.023 & 0.47 & -2.34 & 415\\ 
\enddata
\tablecomments{Simulation parameters are the binary eccentricity $e_{\rm b}$ and the accretion radius $r_{\rm acc}$. The results include
the long-term averages of the binary's angular momentum change rate $\langle \dott{J}_{\rm b}\rangle$, the spin torque 
$\langle \dott{S}_{\rm b}\rangle=\langle \dott{S}_1\rangle+\langle \dott{S}_2\rangle$,
the semimajor axis change rate $\langle \dot{a}_{\rm b}\rangle$ and the eccentricity
change rate $\langle \dot{e}_{\rm b}\rangle$. 
The averaged mass accretion rate onto the binary $\langle \dott{M}_{\rm b}\rangle$ equals to the mass supply rate $\dott{M}_0$.
The binary's {\it orbital} angular momentum change rate is $\langle \dott{L}_{\rm b}\rangle=\langle \dott{J}_{\rm b}\rangle-\langle \dott{S}_{\rm b}\rangle$. Averages are taken over a time interval $T$(chosen to cover one full cycle in alternating accretion).}
\end{deluxetable}
\capstarttrue
%%%%%%%%%%%%%%%%%%%%%%%%%%%%%%%%%%%%%%%%%%%%%%%%%%%%%%%%%%a

The main conclusion drawn from Figure~\ref{fig:angmom_components_examples} is that, provided that steady-state is reached, the net
transfer of angular momentum to the binary $\langle\dott{J}_{\rm b}\rangle$ is positive for various values of $e_{\rm b}$.
As in the circular binary case, to obtain a reliable measurement of $\langle\dott{J}_{\rm b}\rangle$, it is important to capture accurately 
the gravitational torques from the CSDs, which are positive and can counterbalance the negative
torques due to the CBD.

We summarize the results of our simulations in Table~\ref{tab:results}. In all cases explored -- a circular binary with three values of $r_{\rm acc}$
and three eccentric binaries -- the binary experiences a net gain in angular momentum. 
There are multiple consequences of having $\langle\dott{J}_{\rm b}\rangle>0$. In particular, we are interested in translating
$\langle\dott{M}_{\rm b}\rangle$ and $\langle\dott{J}_{\rm b}\rangle$ into secular changes in the orbital
elements of the binary. We explore this in the next section.

%%%%%%%%%%%%%%%%%%%%%%%%%%%%%%%%%%%%%%%%%a
\begin{figure*}[t!]
\centering
\includegraphics[width=0.49\textwidth]{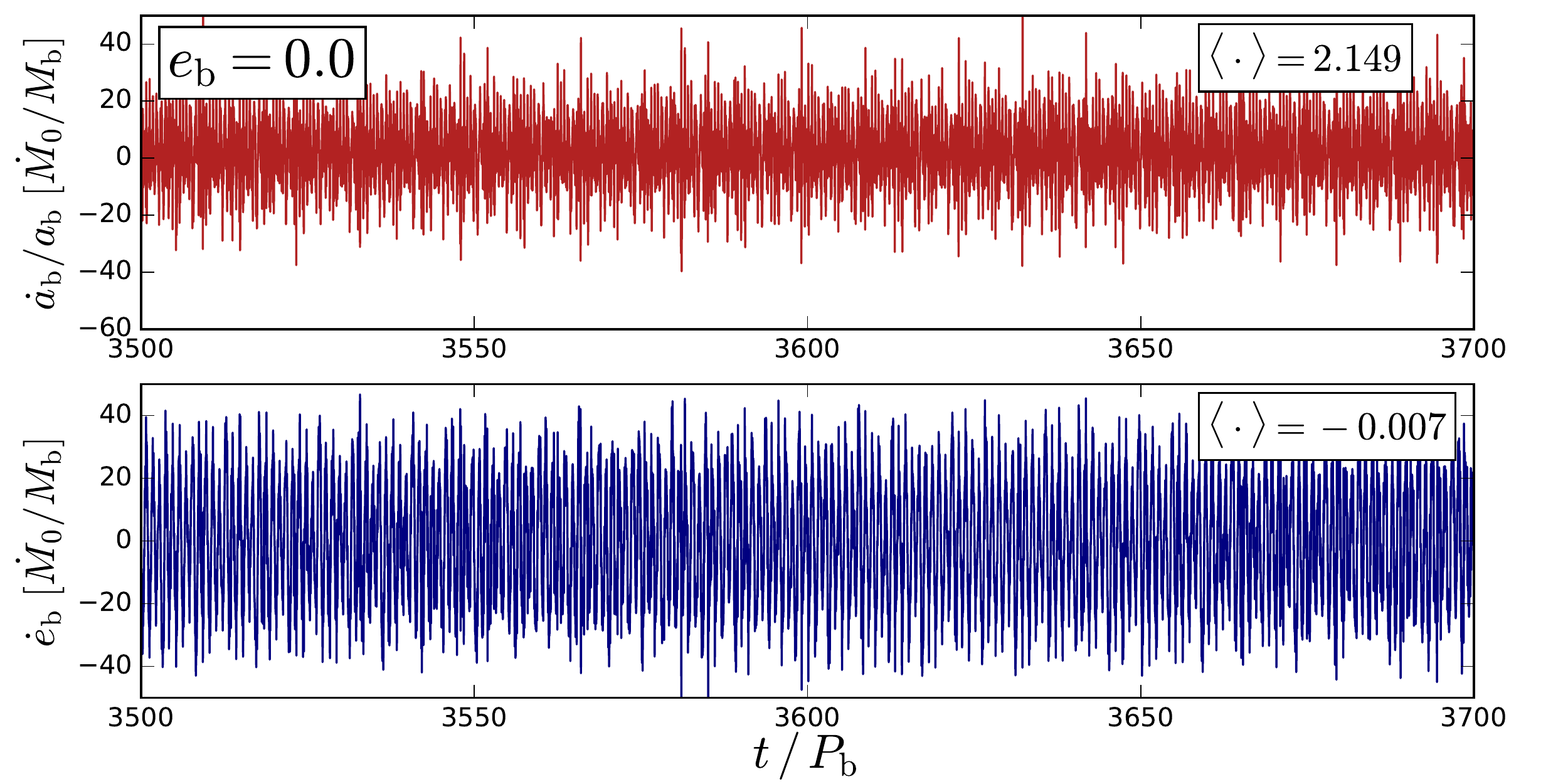}
\includegraphics[width=0.49\textwidth]{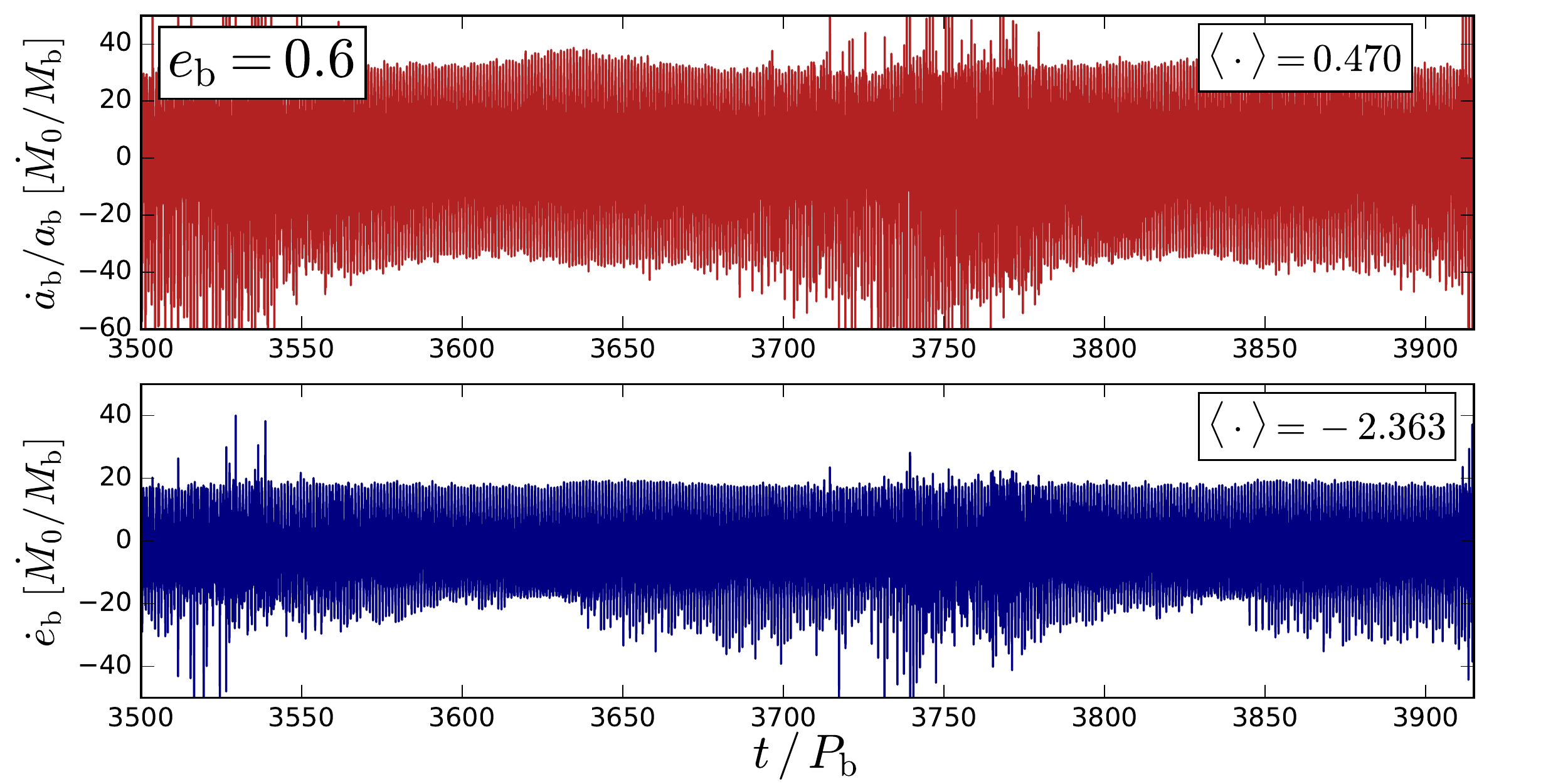}
\caption{Binary semimajor axis change rate (top, red) and the binary eccentricity 
change rate (bottom, blue) for circular and eccentric binaries according to
Eqs.~(\ref{eq:eccentricity_evolution}) and~(\ref{eq:semimajor_evolution}) and the simulation results of Sections~\ref{sec:circular_binary} and~\ref{sec:eccentric_binary}.
Left panels ($e_{\rm b}=0$): the mean values $\langle\dot{a}_{\rm b}\rangle\approx 2.15 a_{\rm b}\dott{M}_0/M_{\rm b}$ and $\langle\dot{e}_{\rm b}\rangle\approx 0$,
indicate that a circular binary tends to remain circular, but that its semimajor axis grows in time. 
Right panels ($e_{\rm b}=0.6$): the binary expands and circularizes ($\langle\dot{a}_{\rm b}\rangle\approx 0.38 a_{\rm b}\dott{M}_0/M_{\rm b}$ and
 $\langle\dot{e}_{\rm b}\rangle\approx -2.34\dott{M}_0/M_{\rm b}$). 
Both circular and eccentric binaries exhibit similar RMS variability in $\dot{a}_{\rm b}$ and  $\dot{e}_{\rm b}$, except during the
the irregular bursts that take place each time $\dot{q}_{\rm b}$ switches signs in the eccentric case (at $t/P_{\rm b}\sim3520$ and $\sim3740$).
\label{fig:orbital_elements}}
\end{figure*}
%%%%%%%%%%%%%%%%%%%%%%%%%%%%%%%%%%%%%%%%%a

\section{Orbital Evolution of Accreting Binaries}\label{sec:orbital_evolution}
For a circular, equal-mass binary, the time-averaged rates  $\langle\dott{M}_{\rm b}\rangle$ and
 $\langle\dott{L}_{\rm b}\rangle\simeq\langle\dott{J}_{\rm b}\rangle$
completely determine the secular semi-major axis evolution  $\langle\dot a_{\rm b}\rangle$ of the binary
 (assuming that the binary remains circular). For an eccentric binary, more information is needed. The orbital evolution 
of variable mass binaries has long been studied \citep[e.g.][]{bro25,jea25,had63,ver72,ver11}. 
\citet{had63} derived the evolution equation for the binary orbital elements due to time-varying
 mass loss by introducing a fictitious force.  Here, we give a much simpler derivation of 
 $\dot a_{\rm b}$ and $\dot e_{\rm b}$ for a mass losing/gaining binary using energy and angular 
 momentum conservation.  We employ two methods to evaluate 
 $\langle\dot a_{\rm b}\rangle$ and $\langle\dot e_{\rm b}\rangle$  using our simulation output.

Consider the specific angular momentum $\mathbf{l}_{\rm b}=\mathbf{r}_{\rm b}\times\dot{\mathbf{r}}_{\rm b}$ and specific
energy $\mathcal{E}_{\rm b}=\tfrac{1}{2}\dot{\mathbf{r}}_{\rm b}^2-\mathcal{G}M_{\rm b}/r_{\rm b}$ of the binary. 
The equation of motion is ${d\dot{\mathbf{r}}_{\rm b}}/{dt}=-({\mathcal{G}{M}_{\rm b}}/{r_{\rm b}^3}) {\mathbf{r}}_{\rm b} +  \mathbf{f}_{\rm ext}$
where $\mathbf{f}_{\rm ext}\equiv \mathbf{f}_{\rm ext,1}- \mathbf{f}_{\rm ext,2}$ is the external force (other than the mutual Keplerian force), 
and $\mathbf{f}_{\rm ext,i}$ is the force per unit mass acting on $M_i$; for accreting binaries,
$\mathbf{f}_{{\rm ext},i}=\mathbf{f}_{{\rm grav},i}+\mathbf{f}_{{\rm acc},i}$  (see Section~\ref{sec:direct_torques}).
The rates of change
in $\mathbf{l}_{\rm b}$ and $\mathcal{E}_{\rm b}$ due to  $\mathbf{f}_{\rm ext}$  and $\dott{M}_{\rm b}$ are
\begin{equation}\label{eq:ldot}
\frac{d\mathbf{l}_{\rm b}}{dt} = \mathbf{r}_{\rm b}\times \mathbf{f}_{{\rm ext}}
\end{equation}
and
\begin{equation}\label{eq:energy_dot}
\begin{split}
\frac{d\mathcal{E}_{\rm b}}{dt} & = -\frac{\mathcal{G}\dot{M}_{\rm b}}{r_{\rm b}}
 +\frac{\mathcal{G}{M}_{\rm b}}{r_{\rm b}^3} {\mathbf{r}}_{\rm b}\cdot \dot{\mathbf{r}}_{\rm b}
 +\dot{\mathbf{r}}_{\rm b}\cdot \frac{d\dot{\mathbf{r}}_{\rm b}}{dt}\\
&=-\frac{\mathcal{G}\dot{M}_{\rm b}}{r_{\rm b}}+
\dot{\mathbf{r}}_{\rm b}\cdot \mathbf{f}_{\rm ext}~~.
\end{split}
\end{equation}
Since  $e_{\rm b}^2=1+2l_{\rm b}^2\mathcal{E}_{\rm b}/(\mathcal{G}M_{\rm b})^2$
and $\mathcal{E}_{\rm b}=-\mathcal{G}M_{\rm b}/(2a_{\rm b})$, then,
\begin{equation}
\label{eq:edot}
-\frac{2e_{\rm b}\dot{e}_{\rm b}}{1-e_{\rm b}^2}=2\frac{\dot{l}_{\rm b}}{l_{\rm b}}+\frac{\dott{\mathcal{E}}_{\rm b}}{\mathcal{E}_{\rm b}}-2\frac{\dot{M}_{\rm b}}{M_{\rm b}}
\end{equation}
and
\begin{equation}\label{eq:adot}
\frac{\dot{a}_{\rm b}}{a_{\rm b}}=-\frac{\dott{\mathcal{E}}_{\rm b}}{\mathcal{E}_{\rm b}}+\frac{\dot{M}_{\rm b}}{M_{\rm b}}~~.
\end{equation}
Applying Equation~(\ref{eq:energy_dot}) to our simulation output (and using the same procedure as described 
in Section~\ref{sec:direct_torques}), we can evaluate $\dot{\cal E}_{\rm b}/\cal E_{\rm b}$ for the accreting binary.
 This is shown, together with $\dott{M}_{\rm b}/M_{\rm b}$ and $\dot l_{\rm b}/l_{\rm b}$, in 
 Figure~\ref{fig:angmom_energy_change} for $e_{\rm b}=0$ (left panels) and $e_{\rm b}=0.6$
(right panels). The linear combination of these quantities gives $d{e}_{\rm b}^2/dt$ (Equation~\ref{eq:edot})
and $\dot{a}_{\rm b}$ (Equation~\ref{eq:adot}). For the $e_{\rm b}=0$ binary, we find
$\langle d{e}_{\rm b}^2/dt\rangle\ll \langle\dott{M}_{\rm b}\rangle/{M}_{\rm b}$ (i.e., it is consistent with zero) 
and $\langle\dot{a}_{\rm b}\rangle/a_{\rm b}\approx 2.15(\dott{M}_0/M_{\rm b}) $. 
This implies that the binary {\it expands} while remaining circular. For the  $e_{\rm b}=0.6$ binary, 
 $\langle d{e}_{\rm b}^2/dt\rangle\approx-2.84(\dott{M}_0/M_{\rm b})$,
 or  $\langle d{e}_{\rm b}/dt\rangle\approx-2.34(\dott{M}_0/M_{\rm b})$, and 
$\langle\dot{a}_{\rm b}\rangle/a_{\rm b}\approx 0.47(\dott{M}_0/M_{\rm b})$. In this case,
the binary still expands -- albeit five times slower than its circular counterpart -- but it tend to circularize while doing so.
These results present a significant challenge to the standard view that binaries surrounded by CBDs 
must shrink (see Section~\ref{sec:discussion}).

We note that Equations~(\ref{eq:edot})-(\ref{eq:adot}) are equivalent to the 
more complex-looking expressions derived by previous authors. 
For example, substituting Eqs.~(\ref{eq:ldot}) and ~(\ref{eq:energy_dot}) into Equation~(\ref{eq:edot}),
and using $\dot{\mathbf{r}}_{\rm b}\cdot\mathbf{f}_{\rm ext}=\Omega_{\rm b}a_{\rm b}(1-e_{\rm b}^2)^{-1/2}[e_{\rm b}\sin\phi f_{{\rm ext},r}+
(1+e_{\rm b}\cos\phi) f_{{\rm ext},\phi}]$ and 
$|\mathbf{r}_{\rm b}\times\mathbf{f}_{\rm ext}|=a_{\rm b}(1-e_{\rm b}^2)(1+e_{\rm b}\cos\phi)^{-1}f_{{\rm ext},\phi}$,
one can write in terms of the true anomaly $\phi$:
\begin{equation}\label{eq:eccentricity_evolution}
\begin{split}
\dot{e}_{\rm b}=&\sqrt{\frac{a_{\rm b}(1-e_{\rm b}^2)}{\mathcal{G}M_{\rm b}}}
\Bigg[\sin\phi\,f_{{\rm ext},r}\\
&~~~~~~~+\Big(\frac{e_{\rm b}+2\cos\phi+e_{\rm b}\cos^2\phi}{1+e_{\rm b}\cos \phi}\Big)f_{{\rm ext},\phi}\Bigg]
-\frac{\dot{M}_{\rm b}}{M_{\rm b}}({e_{\rm b}+\cos \phi})
\end{split}
\end{equation}
and
\begin{equation}\label{eq:semimajor_evolution}
\begin{split}
\dot{a}_{\rm b}=2\sqrt{\frac{a_{\rm b}^3}{\mathcal{G}M_{\rm b}(1-e_{\rm b}^2)}}\bigg[
e_{\rm b}\sin\phi\,f_{{\rm ext},r}+(1+e_{\rm b}\cos\phi)f_{{\rm ext},\phi}\bigg]\\
-a_{\rm b}\frac{\dot{M}_{\rm b}}{M_{\rm b}}\frac{({1+2e_{\rm b}\cos \phi}+e_{\rm b}^2)}{1-e_{\rm b}^2}~.
\end{split}
\end{equation}
When setting $\mathbf{f}_{\rm ext}=0$, we recover the known expressions for the effect of mass loss on $e_{\rm b}$ and $a_{\rm b}$ \citep{had63}.
If $\dott{M}_{\rm b}=0$, we simply recover the perturbation equations for osculating orbital elements \citep{bur76,mur00}.
Furthermore, these expression for $\dot{e}_{\rm b}$ and $\dot{a}_{\rm b}$ are also the coplanar equivalent to those derived by \citet{ver13}
(their Eqs. 29 and 30), in which case the external forces are a result of anisotropic mass loss.

We can also use  Eqs.~(\ref{eq:edot})-(\ref{eq:adot}) 
 to directly compute the instantaneous rates of change $\dot{e}_{\rm b}$ and $\dot{a}_{\rm b}$.
  The results are shown in Figure~\ref{fig:orbital_elements}.
The long-term time-average of these time series provides the secular
evolution of the orbital elements. For the circular binary, we obtain that $\langle\dot{a}_{\rm b}\rangle\approx 2.15a_{\rm b}\dott{M}_0/M_{\rm b}$, and
$\langle \dot{e}_{\rm b}\rangle\approx0$ in agreement with Figure~\ref{fig:angmom_energy_change}. 
For the $e_{\rm b}=0.6$ binary, again we find agreement with Figure~\ref{fig:angmom_energy_change}:
$\langle\dot{a}_{\rm b}\rangle\approx 0.47a_{\rm b}\dott{M}_0/M_{\rm b}$ and $\langle \dot{e}_{\rm b}\rangle\approx-2.34\dott{M}_0/M_{\rm b}$.

Table~\ref{tab:results} includes the measured values of $\langle \dot{a}_{\rm b}\rangle$ and $\langle \dot{e}_{\rm b}\rangle$ for the six simulations presented in this work.
In all configurations explored, $\langle \dot{a}_{\rm b}\rangle>0$, i.e.,  binaries expand as they accrete. 
The behavior of  $\langle \dot{e}_{\rm b}\rangle$ is not monotonic in $e_{\rm b}$:
circular binaries remain circular and high eccentricity binaries circularize; however, the moderate eccentricity  binary ($e_{\rm b}=0.1$) has $\langle \dot{e}_{\rm b}\rangle>0$.  This behavior
in $\langle \dot{e}_{\rm b}\rangle$ is reminiscent of a result reported by \citet{roe11} (simulations of very massive CBD around live binaries), in which a ``fixed point''
for which $\langle \dot{e}_{\rm b}\rangle=0$ can be inferred for some intermediate value of $e_{\rm b}$. 
 {This behavior is perhaps related to the bifurcation in CBD precession rates
measured by \citet{thu17}}.
This
result is intriguing and warrants a more thorough exploration of different values of $e_{\rm b}$.

%%%%%%%%%%%%%%%%%%%%%%%%%%%%%%%%%%%%%%%%%%%%%%%%%%%%%%%%%%%%%%%%%%
\section{Summary and Discussion}\label{sec:discussion}

\subsection{Summary of Key Results}%%%%%%%%%%%%%%%
We have carried out a suite of two-dimensional, viscous hydrodynamical
simulations of circumbinary disk accretion using {\footnotesize AREPO} \citep{spr10a},
 a finite-volume method on a freely moving Voronoi mesh.  As in our previous work
(\citealp{mun16b}; ML16), which also used the {\footnotesize AREPO} code, we can
robustly simulate accretion onto eccentric binaries
without the constraints imposed by structured-grid numerical schemes. Most importantly,
using the {\footnotesize AREPO} code allows us to follow the mass accretion through a wide
radial extent of the circumbinary disk (up to $70a_{\rm b}$, where $a_{\rm b}$ is
the binary semi-major axis), leading to accretion onto the individual
members (``stars'' or ``black holes'') of the binary via circum-single
disks (down to spatial scales of $0.02a_{\rm b}$). 
In ML16, we focused on accretion variability on different timescales. In this paper, we study the 
the angular momentum transfer to the binary, and
for the first time,
determined the long-term evolution rates of orbital elements (semi-major axis and eccentricity) for 
binaries undergoing circumbinary accretion.  This work also goes
significantly beyond our other recent paper (\citealp{mir17};
MML17), which used the polar-grid code {\footnotesize PLUTO} (excising the
innermost ``cavity'' region surrounding the binary) to explore the
dynamics and angular momentum transfer within circumbinary disks.

As in ML16, our simulations focus on the well-controlled numerical
setup where mass is supplied to the outer disk ($R_{\rm out}=70 a_{\rm b}$)
at a fixed rate. We consider equal-mass binaries, use locally
isothermal equation of state (corresponding to $H/R= 0.1$) and
adopt the $\alpha$-viscosity prescription with $\alpha=0.1$.  The
``stars'' or ``black holes'' are treated as absorbing spheres of
radii $r_{\rm acc}\ll a_{\rm b}$ (our canonical value is $r_{\rm
acc}=0.02a_{\rm b}$).  In all our simulations, we evolve the system for a
sufficiently long time so that a quasi-steady state is reached, in
which the time-averaged mass accretion rate onto the binary matches
the mass transfer rate across the disk, and both are equal to
the mass supply rate at $R_{\rm out}$. Our key findings pertain
to the angular momentum transfer and the secular evolution of the 
binary in such a quasi-steady state:
\begin{itemize}
\item[$(i)$] For all cases studied in this paper (with various values of
binary eccentricities; see Table~\ref{tab:results}), the time-averaged angular
momentum transfer rate to the binary is positive 
($\langle \dott{J}_{\rm b}\rangle >0$), i.e., the binary gains angular momentum. We determined
this $\langle \dott{J}_{\rm b}\rangle$ by directly computing the gravitational
torque and accretion torque on the binary.  In addition, we computed
the net angular momentum current (transfer rate) $\dott{J}_{\rm d}$ (Equation~\ref{eq:angular_momentum_balance})
across the circumbinary disk and find that its time average $\langle \dott{J}_{\rm d}\rangle$ is
constant (independent of $R$; see Figs.~\ref{fig:angmom_currents} and~\ref{fig:angmom_currents_ecc}) 
and equal to $\langle \dott{J}_{\rm b}\rangle$.
This agreement shows that global balance is reached 
in our simulations in terms of both mass and angular momentum transfer.

\item[$(ii)$] Our computed angular momentum transfer rate per unit accreted
mass, $\langle \dott{J}_{\rm b}\rangle /\langle \dott{M}_{\rm b}\rangle$, lies
in the range $(0.4-0.8)a_{\rm b}^2\Omega_{\rm b}$ (where $\Omega_{\rm b}$ is the orbital
angular velocity of the binary) and depends on the binary 
eccentricity (see Table~\ref{tab:results}). A small fraction of $\langle \dott{J}_{\rm b}\rangle$
contributes to the spin-up torque on the binary components (``stars'').

\item[$(iii)$]  Using our simulations, we computed the
time-averaged rate of change of the binary orbital energy $\langle
\dot{\cal E}_{\rm b}\rangle$ taking account of gravitational forces and
accretion. Combining $\langle \dot{\cal E}_{\rm b}\rangle$ and $\langle
\dott{J}_{\rm b}\rangle$, we obtain, for the first time,
the secular evolution rates for the binary semi-major axis and eccentricity (Section~\ref{sec:orbital_evolution}).
In all the cases considered, we found that the binary expands while accreting (at a rate of order
$\langle \dot{a}_{\rm b}/{a}_{\rm b}\rangle\sim\langle \dott{M}_{\rm b}\rangle/{M}_{\rm b}$; see Table~\ref{tab:results}).
 Circular binaries remains circular and
high-eccentricity binaries circularize; however, moderate-eccentricity binaries ($e_{\rm b}=0.1$) 
may become more eccentric.

\item[$(iv)$]  We carried out a detailed analysis on the different contributors to the total binary torque
$\dott{J}_{\rm b}$ [see $(i)$]. This analysis showed that while the gravitational torque from 
the circumbinary disk is negative (as expected based on theoretical grounds), the torque
from the inner ``circumstellar'' disks (within $R\lesssim a_{\rm b}$) is overwhelmingly positive,
and is responsible for the overall positive value of $\langle \dott{J}_{\rm b}\rangle$.
A proper calculation of $\langle \dott{J}_{\rm b}\rangle$ requires a careful treatment 
of gas flow within the binary cavity.
\end{itemize}

\subsection{Discussion}%%%%%%%%%%%%%%%%%

\subsubsection{Angular Momentum Transfer: Comparison to Previous Work}%%%%%%%%
\label{sec:previous_work}
The most important result of our study is that in quasi-steady-state,
the binary gains angular momentum. This result was also strongly suggested
in our recent study (MML17) using the polar-grid code {\footnotesize PLUTO}, in which the innermost region of
the circumbinary cavity is excised from the simulation domain. MML17 showed that quasi-steady-state can be reached in the CBD, with global balance of
both $\langle\dott{M}_{\rm d}\rangle(R)$ and $\langle\dott{J}_{\rm d}\rangle(R)$. In fact, for $e_{\rm b}=0$, the accretion eigenvalue 
obtained by MML17  ($l_0\approx0.8a_{\rm b}^2\Omega_{\rm b}$ )
is close to the value obtained in this paper ($l_0\simeq0.68a_{\rm b}^2\Omega_{\rm b}$; Equation~\ref{eq:accretion_eigenvalue}).
At high eccentricities, however, the mismatch between our new results and those of MML17 grows  (e.g., they found $l_0\gtrsim a_{\rm b}^2\Omega_{\rm b}$
for $e_{\rm b}=0.6$).  This discrepancy may be understood from the fact that for finite 
 eccentricities, the mass flow $\dott{M}_{\rm d}(R)$ at $R=a_{\rm b}(1+e_{\rm b})$
(the inner boundary of the computation domain adopted by MML17) can be both inward and outward (see Fig.~2 of ML16), a 
behavior that a diode-like open boundary (adopted by MML17) cannot capture. 
The mismatch also underscores the importance of resolving the flows inside the cavity. Because 
the simulations of MML17 did not fully include the inner cavity,
one should take those results as suggestive. In this paper, calculating
$\langle\dott{J}_{\rm b}\rangle$ in two different ways and obtaining
 agreement (direct torque and angular momentum current across the CBD), we have obtained
robust evidence that the binary can gain angular momentum while accreting ($\langle\dott{J}_{\rm b}\rangle>0$). This result is insensitive to the 
accretion radius (the size of the "stars") 
as long as  $r_{\rm acc}\lesssim 0.1 a_{\rm b}$ (see Table~\ref{tab:results}).

Our result ($\langle \dot J_{\rm b}\rangle >0$) contradicts the widely held
view that a binary loses angular momentum to its surrounding disk
and experiences orbital decay. In fact, while there have been many 
numerical simulations of circumbinary accretion (see references in the Section~\ref{sec:introduction}), 
only a few address the issue of angular momentum transfer 
in a quantitative way. Three such studies use simulations that excise the inner cavity
(\citealp{mac08,shi12}; MML17).
The work of \citet{mac08} was the most similar to 
MML17: these authors considered $H/R = 0.1$, a disk viscosity with
$\alpha= 0.01$, and adopted 
a polar grid in the domain between $R_{\rm in} = a_{\rm b}$ and $R_{\rm out} = 100a_{\rm b}$.
They found a reduction of mass accretion onto the
binary and the dominance of the gravitational torque relative
to advective torque (therefore a negative net torque on the binary).
However, with their small viscosity parameter (by contrast, MML17 used 
$\alpha=0.1$ and 0.05), the ``viscous relaxation'' radius
$t = 4000P_{\rm b}$ (the typical duration of their runs) is only 
about $3a_{\rm b}$, and their surface density profile is far from
steady state even for $R\gg R_{\rm in}$. We suggest that the result of
\citet{mac08} reflects a ``transient'' phase of their simulations.
\citet{shi12} obtained a positive value of $\dott{J}_{\rm b}$
in their 3D MHD simulations of circumbinary disks (truncated at 
$R_{\rm in}=0.8a_{\rm b}$). However, the duration
of their simulations is only $\sim100P_{\rm b}$, and it is unlikely that a quasi-steady
state is reached. Their value of $l_0$, which is too small to cause orbital
expansion,  may not properly characterize the long-term evolution of the binary.
As noted above, our previous study (MML17) 
using the polar-grid code {\footnotesize PLUTO} (with $R_{\rm in}=a_{\rm b}(1+e_{\rm b})$ for eccentric
binaries) did achieve quasi-steady-state in terms of mass and angular momentum
transport in the CBD. For $e_{\rm b}=0$, the value of $l_0$ measured by MML17 was close to the value presented in this paper (Table~\ref{tab:results}), while the 
$l_0$ values for $e_{\rm b}>0$ were unreliable for understandable reasons (see above).

In a recent paper, \citet{tan17} (TMH17) carried out hydrodynamical simulations
of accretion onto circular binaries using the {\footnotesize DISCO} code \citep{duf12,duf16}.  Their
simulations are similar to ours in many respects (i.e., locally
isothermal EOS with $H/R=0.1$, $\alpha=0.1$). Their accretion
presciption is different from ours: inside a ``sink'' radius (measured
from each ``star''), they assumed that the gas is depleted at the rate
$d\Sigma/dt =-\Sigma/t_{\rm sink}$, with $r_{\rm sink}=0.1a_{\rm b}$ and
$t_{\rm sink}$ a free parameter. By contrast, the {\footnotesize AREPO} code in its Navier-Stokes formulation \citep{mun13a}
self-consistently incorporates viscous transport in all of our
computation domain in a unified manner, and gas accretion onto the
``stars'' is treated using an absorbing sink algorithm with a single parameter $r_{\rm
acc}=0.02a_{\rm b}$ (see Section~\ref{sec:num_methods}).
Using a direct force computation, TMH17 found that the
net torque on the binary is negative unless $t_{\rm sink}$ is much
smaller than $P_{\rm b}$. 
The discrepancy between our results and those of TMH17 is important and deserves
further study/comparison. For now, we offer the following comments: 
($i$) In our simulations, we have achieved global mass transfer balance
($\langle\dott{M}_1\rangle +\langle\dott{M}_2\rangle \simeq \langle \dott{M}_{\rm d}\rangle(R)
\simeq\dott{M}_0$; see Figs.~\ref{fig:accretion_rates} and~\ref{fig:angmom_components}) and global angular momentum transfer balance
($\langle\dott{J}_{\rm b}\rangle\simeq \langle\dott{J}_{\rm d}\rangle(R)\simeq{\rm constant}$; see Figs.~\ref{fig:angmom_currents} and~\ref{fig:angmom_currents_ecc}); 
TMH17 did not demonstrate that they have achieved these. 
($ii$) The mass depletion prescription adopted by TMH17 may be problematic in terms
of mass conservation. TMH17 found that the more ``filled'' the CSDs are
(corresponding to larger $t_{\rm sink}$), the more negative the net torque
on the binary is (see their Fig.~2); this is counterintuitive, as on physical grounds
one expects that CSDs contribute a positive gravitational torque on the binary \citep{sav94}.
($iii$) It is unclear whether the definition and calculation of the
accretion torque in TMH17 are consistent with angular momentum conservation
(see their Eq.~8).  
%By contrast, our calculations of $\dott{J}_{\rm d,grav}$ -- as depicted in Figs.~\ref{fig:angmom_currents} and~\ref{fig:angmom_currents_ecc} -- 
%show that this quantity is significant only for $R\lesssim 3a_{\rm b}$.

Turning to eccentric binaries, our paper represents the first work to compute
the {\it secular} evolution rates of $a_{\rm b}$ and $e_{\rm b}$, consistently taking into account accretion
via direct, long-term simulations.
Most previous (analytic) works
\citep[such as][]{art96,lub00b} assumed that accretion was negligible ($\dott{M}_{\rm b}=0$) and
that $\dot{a}_{\rm b}$ and $\dot{e}_{\rm b}$ were solely due to gravitational Lindblad torques.
 {\citet{hay09} carried out an analytic calculation including the effect of accretion coupled to
the Lindblad torques from the CBD; in contrast to our new findings,
he concluded that, for $\alpha=h_0=0.1$, binaries migrate inward if they accrete at the same rate 
as the CBD.}

\subsubsection{Caveats and Limitations of our work}%%%%%%%%%%%%%
In this paper, we have ``solved'' the  problem of circumbinary accretion in the ``cleanest''  scenario and in quasi-steady-state, taking advantage of {\footnotesize AREPO}'s
capability to resolve a wide dynamical range in space (from $70a_{\rm b}$ down to $0.02a_{\rm b}$). We should, however,
bear in mind the limitations of our work when confronting with 
real astrophysical applications.  

Our simulations are two-dimensional and adopt a simple equation of state as well as the $\alpha$-viscosity prescription.
Three-dimensional MHD simulations \citep{shi12,nob12,shi15} would represent
the transport of angular momentum more realistically. 
In addition, general
relativistic effects are bound to be important at close separations in the case of SMBBHs \citep{far12}. 

 {The value of $\alpha$ chosen (0.1) in our simulations is in part due to computational expediency. 
A larger viscosity results in a shorter viscous time  -- still much longer than all other relevant timescales -- which allows us
to reach the quasi-steady state in the inner region ($R\lesssim 10a_{\rm b}$) of the CBD.
Very small values of $\alpha$ would
 prevent us from attaining such steady-state.
MML17 considered $\alpha=0.1$ and $0.05$ in their 
{\footnotesize PLUTO} simulations and found the same $l_0$ value for circular binaries.
At very low viscosities, one can expect more pronounced spiral
arms (in the CBD and CSDs) which would increase the tidal torques; at the same time, the tidal cavity surrounding the binary becomes larger
and the CSDs are truncated at smaller radii \citep{art94,mir15}, thus removing the gas that would exert the strongest torques.
These opposing effects may cancel each other out (at least for $\alpha=0.05-0.1$ and $e_{\rm b}=0$) as suggested
by the results of MML17. On the other hand, the SPH simulations by \citet{rag16} suggest
 that accretion may be suppressed when the viscosity is sufficiently small (small $\alpha$ or small $h_0$), although it
 is not clear if those simulations were run for a sufficiently long integration time.
 Exploring the dependence of our results on $\alpha$ and $h_0$ will be the subject of
future work.
}

We have assumed a steady mass supply at large distance, which may not exist under realistic astrophysical conditions.
We expect that our general results should hold even if the external supply rate $\dott{M}_0$ varies in time, provided it does so smoothly and slowly,
on a timescale longer than the accretion time at the outer edge of the CBD, $\tau_{\rm acc}\sim R_{\rm out}^2/\nu$.
As several of the quantities measured in this work
are scaled by $\dott{M}_0$, such as $\dott{J}_{\rm b}$, $\dot{a}_{\rm b}$ and $\dot{e}_{\rm b}$, a slow change of the supply rate will translate into a proportional
amplification or reduction of these quantities, without affecting their qualitative behavior. On the other hand,
 clear violations of the steady-supply condition -- such as accretion via gas clumps  \citep[e.g.][]{goi16}, or from tori of small radial extent -- might change 
the response of the binary.

We have focused on $q_{\rm b}=1$ in this paper, but the behavior of $q_{\rm b}<1$ binaries may be different.
In particular, the phenomenon of alternating accretion described in ML16, attributed to apsidal precession in the CBD, has not been thoroughly explored
for mass ratios different from unity.  The value of $\langle \dot{q}_{\rm b}\rangle $ -- and whether it averages to zero over secular timescales -- 
is very important for properly determining $\langle\dott{J}_{\rm b}\rangle$ when $q_{\rm b}\neq1$ (Eqs.~\ref{eq:angmom_change} and~\ref{eq:reduced_mass_change}).
The sign of $\langle\dot{q}_{\rm b}\rangle$ from CBD accretion is still an unresolved question, and
recent work by \citet{you15} suggests that it depends on the aspect ratio of the disk \citep[see also][]{cla08}. No grid-based simulation has explored the sign of
$\langle\dot{q}_{\rm b}\rangle$ while systematically varying both $q_{\rm b}$ and $e_{\rm b}$, and this will be the subject of future work.

Finally, we do not include self-gravity of the disk in our simulations, as we have implicitly assumed that the local disk mass,
 $R^2 \Sigma (R)$ (with $R\sim 5 a_{\rm b}$),  is much less than the binary mass.  
 A massive, self-gravitating disk may qualitatively change the dynamics of circumbinary accretion studied in this
  paper \citep[e.g.,][]{cua09,roe12},  {although some disk fragmentation studies have already 
  reported outward migration of proto-binaries \citep{kra10a}}; see below.

\subsubsection{Implications for Disk-Assisted Binary Decay}%%%%%%%%%%%%
\paragraph{Supermassive Binary BHs}%%%%%%%%%%%%%%%%%%
 It is widely assumed that circumbinary disks around SMBBH extract angular momentum from such binaries, shrinking their orbital radii from 
 $\sim1$ pc to $\sim0.01$ pc, before gravitational radiation 
can push the binary toward merger within a Hubble time \citep[e.g.][]{beg80,arm02,mac08,raf16,tan17}.
 Our findings contradict this assumption, thus putting into question
the relevance of circumbinary gas as a solution to ``final parsec problem''. 
At least for equal-mass binary black holes and non-self-gravitating disks, our results suggest that circumbinary accretion may prevent
rather than assist the merger of such objects.

\paragraph{Formation of stellar Binaries}%%%%%%%%%%%%%%%%%%
The finding that binaries embedded in accretion disks are able to gain angular momentum poses
problems to binary formation theory. Of the proposed mechanisms for binary formation at intermediate separations \citep[$\lesssim10$~au; e.g.,][]{toh02}
the still qualitative ``fragmentation-then-migration'' scenario has emerged as a strong contender over the past decade \citep[e.g.,][]{kru09,kra10a,zhu12}. This mechanism
is grounded on the difficulty for massive disks to fragment at distances smaller than $50-100$~AU from the central object \citep{mat05,raf05,bol06,whi06,sta08,cos09,kra10b},
and thus requires binaries to reduce their separations by one or two orders of magnitude after they have formed \citep[see][for a recent analysis of binaries at small separations and their 
possible formation mechanisms]{moe18}.
If binaries embedded in disks tend to expand, the viability of this mechanism would be questionable \citep[e.g.,][]{sat17}.

\paragraph{\it How can binaries still shrink?}%%%%%%%%%%%%%%%%%%
Given the idealized nature of our simulations, we cannot rule out the possibility 
of binary shrinkage. We envision several possibilities where binary decay may occur:

($i$) Unequal-mass binaries and locally massive disks. We have only explored binaries
with $q_{\rm b}=1$ and small disk masses, but systems with $q_{\rm b}<1$ and disks with local mass $\sim\Sigma a_{\rm b}^2 \gtrsim M_2$
could behave very differently \citep{cua09,roe12}. 
In particular, when $q_{\rm b}\ll1$, the binary would evolve in a similar fashion as Type II migration \citep[e.g.][]{lin86b,nel00} or a modified version
of it \citep[e.g.,][]{duf14,dur15}. Such massive disks may exist in the early stage immediately following galaxy merger (producing a massive gas torus surrounding a SMBH binary) or the
early  (class zero) stage of stellar binary formation.

($ii$) Three-dimensional angular momentum loss effects. Our two-dimensional calculations cannot capture effects that
are intrinsically 3D, such as winds and outflows. MHD winds can take away angular momentum in accretion disks, and
such a mechanism may operate in in circumbinary accretion as well. The 
3D ideal MHD simulations of \citet{kur17} suggest that newly formed binaries could contract significantly within hundreds of orbits as they grow in mass by a factor of $\sim100$,
again implying that binary shrinkage may take place at a much earlier phase of binary formation than the that explored in the present work.

%%%%%%%%%%%%%%%%%%%%%%%%%%%%%%%%%%%%%%%%%%%%%%%%%%%%%%%%%%%%%%%%%%
\acknowledgements
D.J.M. thanks Volker Springel for making {\footnotesize AREPO} available for use in this work. 
We thank Kaitlin Kratter, Yoram Lithwick, Zsolt Reg\'aly, Pawel Artymowicz and Zolt\'an Haiman for useful discussion
and Luke Zoltan Kelley for comments on the manuscript.
This work is supported in part by the NSF grant AST1715246 and NASA grant NNX14AP31G.
This research was supported in part through the computational resources and staff contributions provided for the
 Quest high performance computing facility at Northwestern University which is jointly supported by the Office of the Provost, 
 the Office for Research, and Northwestern University Information Technology.

%%%%%%%%%%%%%%%%%%%%%%%%%%%%%%%%%%%%%%%%%%%%%%%%%%%%%
%%%%%%%%%%%%%%%%%%%%%%%%%%%%%%%%%%%%%%%%%%%%%%%%%%%%%

\acknowledgments{ }

\bibliographystyle{apj}
%\bibliography{disks_bib,hydro_bib,youngstars_bib,dynamics_bib,tides_bib,statistics_bib}

\end{document}